\documentclass[sn-nature]{sn-jnl}

\usepackage{graphicx}%
\usepackage{multirow}%
\usepackage{amsmath,amssymb,amsfonts}%
\usepackage{amsthm}%
\usepackage{mathrsfs}%
\usepackage[title]{appendix}%
\usepackage{xcolor}%
\usepackage{textcomp}%
\usepackage{manyfoot}%
\usepackage{booktabs}%
\usepackage{algorithm}%
\usepackage{algorithmicx}%
\usepackage{algpseudocode}%
\usepackage{listings}%
\usepackage{caption}
\usepackage{subcaption}
\usepackage{color,soul}

\theoremstyle{thmstyleone}%
\theoremstyle{thmstyletwo}%

\theoremstyle{thmstylethree}%

\newcommand{\araa}{Annu. Rev. Astron. Astrophys.}   
\newcommand{\aj}{Astron. J.}   
\newcommand{\apj}{Astrophys. J.}   
\newcommand{\apjl}{Astrophys. J. Lett.}   
\newcommand{\aap}{Astron. Astrophys.}   
\newcommand{\mnras}{Mon. Not. R. Astron. Soc.}   
\newcommand{\nastro}{Nat. Astron.} 

\newcommand{\gaia}{{\it Gaia}}
\newcommand{\bprp}{$BP-RP$}
\newcommand{\ds}{$\Delta s$}
\newcommand{\dv}{$\Delta v$}
\newcommand{\kms}{km s$^{-1}$}
\newcommand{\teff}{$T_{\rm eff}$}
\newcommand{\logg}{$\log g$}
\newcommand{\vt}{$\xi_t$}
\newcommand{\fei}{Fe\,{\sc i}}
\newcommand{\feii}{Fe\,{\sc ii}}
\newcommand{\tcond}{$T_{\rm cond}$}

\newcommand{\mearth}{M$_{\oplus}$}

\newcommand{\mcz}{M$_{\rm CZ}$}

\begin{document}

\title[Article Title]{\bf At least one in a dozen stars exhibits evidence of planetary ingestion\footnote{This paper includes data gathered with the 6.5 meter Magellan Telescopes located at Las Campanas Observatory, Chile. Some of the data presented herein were obtained at the W. M. Keck Observatory, which is operated as a scientific partnership among the California Institute of Technology, the University of California and the National Aeronautics and Space Administration. The Observatory was made possible by the generous financial support of the W. M. Keck Foundation. Based on observations collected at the European Organisation for Astronomical Research in the Southern Hemisphere under ESO programme 108.22EC.001. *E-mail: fan.liu@monash.edu}} 


\author[1,2,3*]{\fnm{Fan} \sur{Liu}}

\author[3,4,5,6,7,8]{\fnm{Yuan-Sen} \sur{Ting}}

\author[3,4]{\fnm{David} \sur{Yong}}

\author[9,10]{\fnm{Bertram} \sur{Bitsch}}

\author[1,3]{\fnm{Amanda} \sur{Karakas}}

\author[2]{\fnm{Michael~T.} \sur{Murphy}}

\author[11,12]{\fnm{Meridith} \sur{Joyce}}

\author[13]{\fnm{Aaron} \sur{Dotter}}

\author[14,15]{\fnm{Fei} \sur{Dai}}

\affil*[1]{\orgdiv{School of Physics and Astronomy}, \orgname{Monash University}, \orgaddress{\city{Clayton}, \postcode{Victoria 3800}, \country{Australia}}}

\affil[2]{\orgdiv{Centre for Astrophysics and Supercomputing}, \orgname{Swinburne University of Technology}, \orgaddress{\city{Hawthorn}, \postcode{Victoria 3122}, \country{Australia}}}

\affil[3]{\orgname{ARC Centre for All Sky Astrophysics in 3D (ASTRO-3D)}, \orgaddress{\city{Canberra}, \state{ACT}, \country{Australia}}}

\affil[4]{\orgdiv{Research School of Astronomy and Astrophysics}, \orgname{Australian National University}, \orgaddress{\city{Weston}, \postcode{ACT 2611}, \country{Australia}}}

\affil[5]{\orgdiv{School of Computing}, \orgname{Australian National University}, \orgaddress{\city{Acton}, \postcode{ACT 2601}, \country{Australia}}}

\affil[6]{\orgdiv{Department of Astronomy}, \orgname{The Ohio State University}, \orgaddress{\city{Columbus}, \postcode{OH 45701}, \country{USA}}}

\affil[7]{\orgdiv{Center for Cosmology and AstroParticle Physics (CCAPP)}, \orgname{The Ohio State University}, \orgaddress{\city{Columbus}, \postcode{OH 43210}, \country{USA}}}

\affil[8]{\orgname{Observatories of the Carnegie Institution of Washington}, \orgaddress{\street{813 Santa Barbara Street}, \city{Pasadena}, \postcode{CA 91101}, \country{USA}}}

\affil[9]{\orgname{Max-Planck-Institut f\"ur Astronomie}, \orgaddress{\street{Knigstuhl 17}, \city{Heidelberg}, \postcode{69117}, \country{Germany}}}

\affil[10]{\orgdiv{Department of Physics}, \orgname{University College Cork}, \orgaddress{\city{Cork}, \postcode{T12 R229}, \country{Ireland}}}

\affil[11]{\orgdiv{HUN-REN Research Centre for Astronomy and Earth Sciences}, \orgname{Konkoly Observatory}, \orgaddress{\street{Konkoly Thege Mikl\'os \'ut 15-17}, \city{Budapest}, \postcode{H-1121}, \country{Hungary}}}

\affil[12]{\orgdiv{CSFK}, \orgname{MTA Centre of Excellence}, \orgaddress{\street{Konkoly Thege Mikl\'os \'ut 15-17}, \city{Budapest}, \postcode{H-1121}, \country{Hungary}}}

\affil[13]{\orgdiv{Department of Physics and Astronomy}, \orgname{Dartmouth College}, \orgaddress{\street{6127 Wilder Laboratory}, \city{Hanover}, \postcode{NH 03755}, \country{USA}}}

\affil[14]{\orgdiv{Division of Geological and Planetary Sciences}, \orgname{California Institute of Technology}, \orgaddress{\street{1200 E California Blvd}, \city{Pasadena}, \postcode{CA 91125}, \country{USA}}}

\affil[15]{\orgdiv{Department of Astronomy}, \orgname{California Institute of Technology}, \orgaddress{\city{Pasadena}, \postcode{CA 91125}, \country{USA}}}



\maketitle

\textbf{Stellar chemical compositions can be altered by ingestion of planetary material \citep{pin01,hb23} and/or planet formation which removes refractory material from the proto-stellar disc \citep{mel09,bo20}. These ``planet signatures" appear as correlations between elemental abundance differences and the dust condensation temperature (\tcond; \citealp{lod03,mel09,cha10}). 
Detecting these planet signatures, however, is challenging due to unknown occurrence rates, small amplitudes, and heterogeneous star samples with large differences in stellar ages \citep{adi14,nis15}, and therefore stars born together (i.e., co-natal) with identical compositions can facilitate such detections. While previous spectroscopic studies were limited to small number of binary stars \citep{ram15,saf17,oh18,nag20,gal21}, the Gaia satellite \citep{gai21} provides new opportunities for detecting stellar chemical signatures of planets among co-moving pairs of stars confirmed to be co-natal \citep{kam19,nel21}. Here we report high-precision chemical abundances for a homogeneous sample of 91 co-natal pairs of stars with a well-defined selection function and identify at least seven new instances of planetary ingestion, corresponding to an occurrence rate of 8\%. An independent Bayesian indicator is deployed, which can effectively disentangle the planet signatures from other factors, such as random abundance variation and atomic diffusion \citep{dot17}. Our study provides new evidence of planet signatures and facilitates a deeper understanding of the star-planet-chemistry connection by providing new observational constraints on the mechanisms of planet engulfment, formation and evolution. 
} 

\bigskip
Using high-precision astrometric data from the Gaia satellite \citep{gai21}, we establish a large and homogeneous new sample of 125 co-moving pairs of stars with a well-defined and unbiased selection function for high-precision spectroscopic analysis \citep{yon23}, aiming to examine these potential planet signatures. 91 pairs of stars with spatial separations $\Delta s < 10^6$\,AU can be regarded as close and co-natal with a shared origin \citep{kam19,nel21}, while another 34 highly-separated (i.e., ``far") co-moving pairs with larger $\Delta s$ ($\geq 10^6$\,AU) are considered as a control sample. High-resolution (R = $\lambda$/$\Delta\lambda$ $\sim$ 50,000--110,000) and high signal-to-noise ratio (S/N $\approx$ 250 per pixel) spectra were obtained from the European Southern Observatory’s Very Large Telescope, the Magellan Telescope, and the Keck Telescope. The precise (differential) stellar parameters of our sample stars, mainly late-F and G dwarfs, were determined \citep{yon23}. 
By employing a strictly line-by-line differential analysis (consistent with parameter determination) that significantly reduces systematic uncertainties, extremely high precision was achieved with $\sim$\,0.015 dex (3.5\%) relative abundance errors for a suite of 21 elements from C to Ce covering a range of \tcond\ values and nucleosynthesis processes (see Methods). Our sample size is about 5\,--\,10 times larger than the previous studies with comparable abundance precision for at least 15 elements (e.g., \citealp{nag20,beh23}). 

\smallskip

The primary approach involved developing and applying Bayesian analysis to our precise abundance data, in conjunction with a model for planetary ingestion. Here we assume that the observed abundance differences result from ingestion of planetary material, and quantitatively examine the mass of bulk Earth material (M$_{\rm E}$) required to match the abundance pattern for a given co-moving pair, along with the corresponding Bayesian evidence ln(Z)$_{\rm planet}$ for the planetary ingestion model. We then compared the results against the flat model (null hypothesis) and atomic diffusion model for each co-moving pair (see Methods). The difference in Bayesian evidence $\Delta$ln(Z) between the planetary ingestion and flat models provides us with a stringent indicator to identify and validate the chemical signatures of ingestion of planetary material. 

The fitting results for an example pair HD\,185726/185689 (Pair 124) are shown in Figure \ref{fig:fitting_data}\,\textbf{a}, which has amongst the largest $\Delta$ln(Z). For this pair, the abundance data agree well with the predicted model of planetary ingestion, but clearly deviate from the flat model and the atomic diffusion model. The distributions of ln(Z)$_{\rm planet}$ and $\Delta$ln(Z) between the planetary ingestion and flat models are shown in Figure \ref{fig:hist_dlnZ}\,\textbf{a}, which are distinguishable between the close and far co-moving pairs, with the far pairs exhibiting smaller $\Delta$ln(Z). Similar tests were applied to a mock noise sample and a mock signal sample (see Methods), demonstrating that we can effectively disentangle the signatures of planetary ingestion from pure measurement noise using $\Delta$ln(Z) as an indicator. A cutoff value of 3.5 for $\Delta$ln(Z) was adopted by inspection of the distributions of the control sample of far co-moving pairs, the mock noise and mock signal samples. We note there are 11 close co-moving (i.e., co-natal) pairs with $\Delta$ln(Z) $>$ 3.5, making them candidates for chemical signatures of ingestion of planetary material.

Furthermore, the Bayesian modelling allowed us to detect, if present, the signatures of atomic diffusion among our sample stars. The {\sc MIST} atomic diffusion models \citep{dot17} include predictions of abundance changes at different stellar evolutionary phases for eight elements. The Bayesian evidence for the planetary ingestion and atomic diffusion models can be calculated for these elements (see Methods). Figure \ref{fig:hist_dlnZ}\,\textbf{b} shows the distributions of differences in Bayesian evidence $\Delta$ln(Z)$_{\rm atom}$ between the planetary ingestion and atomic diffusion models. We found that more than half of the far co-moving pairs have negative values of $\Delta$ln(Z)$_{\rm atom}$ and all of them have $\Delta$ln(Z)$_{\rm atom}$ less than 2.5, demonstrating that the observed abundance differences within some of these pairs are potentially affected by atomic diffusion. In contrast, among the 11 co-moving pairs with the strongest Bayesian evidence for planetary ingestion ($\Delta$ln(Z) $>$ 3.5), 10 have prominent, positive differences in Bayesian evidence between the planetary ingestion and atomic diffusion models ($\Delta$ln(Z)$_{\rm atom}$ $>$ 2.5), indicating that the effects of atomic diffusion can not explain the observed pattern of abundance differences in these co-natal pairs of stars. 

We then examined the trend between abundance differences within a given pair of stars and \tcond\ of 21 elements \citep{lod03} by applying a linear least squares fit to the abundance data (weighted with corresponding uncertainties) for each pair (see Methods for details). In Figure \ref{fig:fitting_data}\,\textbf{b} we present the differential elemental abundances as a function of \tcond\ for an example co-moving pair of stars: HD\,185726/185689 (Pair 124), which exhibits one of the most significant and largest amplitudes in the \tcond\ trends, further confirming our detection. Figure \ref{fig:tcond_slope}\,\textbf{a} shows the spatial separation (\ds) versus the velocity separation (\dv), with the symbol size scaled by the absolute value of \tcond\ slopes. A value of \ds\ = 10$^6$\,AU separates the 91 close and 34 far co-moving pairs. No dependence was found between the \tcond\ slopes and \ds\ for the co-natal sample. Recall that the close pairs are assumed to be co-natal and share the same chemical composition, while the far pairs are expected to show random abundance variations due to different birth environments and/or atomic diffusion as introduced above. Unlike our Bayesian approach, the distributions of \tcond\ trends for the close and far co-moving pairs are almost indistinguishable (especially for those with strong \tcond\ trends) in Figure \ref{fig:tcond_slope}\,\textbf{b}, demonstrating that, as an indicator, \tcond\ trends alone cannot effectively disentangle planet signatures from other effects (e.g., inhomogeneous ISM and/or atomic diffusion). Therefore relying solely on \tcond\ trends may lead to false positives and an overestimation of the occurrence rate of planetary ingestion. 

In this study, we mainly focus on the 91 close, co-natal pairs because the co-moving pairs in our control sample are widely separated, and they may not necessarily have been born with the same initial composition. By assessing the distributions of different indicators, we adopted below criteria as strong evidence of ingestion of planetary material within a given pair: \\ 
(i) $|$\tcond\ slope$| > 4 \times 10^5$ dex\,K$^{-1}$ and significance level $>$ 3\,$\sigma$; \\
(ii) $\Delta$ln(Z) $>$ 3.5; \\
(iii) $\Delta$ln(Z)$_{\rm atom}$ $>$ 2.5. \\ 
Among our co-natal sample, we found that 21 pairs fulfil criterion (i), 11 pairs fulfil criterion (ii), and finally 7 pairs (out of 91) fulfil all of the above criteria (see Bayesian fitting results in Extended Figures \ref{fig:fitting_A5}\,--\,\ref{fig:fitting_A7}). 
We report at least seven new instances of planetary ingestion from our sample of these co-natal pairs, corresponding to an occurrence rate of 8\% ($\pm$3\%, if assuming a Poisson noise). We show that comparing the Bayesian evidence between the planetary ingestion and other models can effectively distinguish the potential planet signatures, leading to stringent discoveries. Previous high-precision studies of binary systems reported in total 7--8 likely instances of planetary ingestion (primarily based upon \tcond\ trends). Our study doubles the number of known instances of planetary ingestion, expanding our knowledge of the chemistry connection between stars and planets.

 The occurrence rate found in this study (8\%) is comparable to the estimation based on the solar twins (15$\pm$9\%; \citealp{ram09}) and is aligned with theoretical predictions from N-body simulations \citep{bi23}. Our value is lower than that in previous studies \citep{spi21}, which presented compiled abundance results for 107 binary systems with precision ranging from 0.02 to 0.1 dex. Among these, 33 pairs exhibited anomalous iron abundance differences, which were attributed to planetary ingestion occurring in 20--35\% of Sun-like stars. When applying the same definition, our results are generally consistent with theirs in terms of the fraction of chemical anomalies \citep{yon23}. The differences in ingestion rates found in this study may be due to a more homogeneous sample and analysis methods that utilize a larger set of elements, as opposed to mainly carbon and iron, in the search for potential planet signatures. 
 Furthermore, our occurrence rate is marginally higher than the prediction ($<$\,4.9\%) from a recent study \citep{beh23}. They attributed the lower occurrence rate to efficient thermohaline mixing that can obscure planet signatures imprinted. However, thermohaline mixing is recognized as being poorly constrained in 1D stellar models \citep{tj22}. Hydrodynamical simulations suggest low thermohaling efficiency in stellar environments \citep{tra11,bro13}, with no consensus on how other instabilities (e.g., caused by rotation) affect thermohaling mixing \citep{tj22}. 
 
Our results (e.g., the observed abundance differences, the reported new candidates with planet signatures, and the occurrence rate) have far-reaching implications for the scenarios of planet formation and engulfment. In the context of planet engulfment, N-body simulations of super-Earth formation \citep{izi21,bi23} show that scattering events mostly happen within the first 100 Myr after system formation. During this time planets are also regularly engulfed by the central star. However, these early engulfment might not be traced by abundance differences after a few Gyr of evolution \citep{beh23}. Therefore we may be witnessing late accretion events caused by outer perturbers (e.g., outer cold giants or stellar flybys) and/or the slow erosion of inner planets' atmospheres, leading to late instabilities \citep{mo20}. This challenges the previously perceived stability of super-Earth systems. Approximately 30--50\% of solar-like stars host inner super-Earths in the range of a few Earth masses \citep{fre13,mul18}. When considering planetary occurrence rates and our ingestion statistics, it suggests that roughly every 4\,--\,10 super-Earth systems could experience late ingestion events. This study thus provides new insights into the long-term evolution of planetary systems.

On the other hand, our identified co-moving pairs showing clear planet signatures can also be interpreted with other planet-related scenarios, such as (a) the formation and accretion of proto-planetary discs \citep{hb23}; (b) the formation of terrestrial planets that remove refractory material from the proto-planetary discs \citep{mel09}; (c) the formation of distant giant planets that prevent the accretion of refractory material from outer regions \citep{bo20}. The above processes are likely to imprint smaller amounts of abundance differences (0.02–0.04 dex; 5–10\%) when compared to that from the later accretion/ingestion events, due to different time-scales and sizes of stellar convection zone (see Methods). 
The amplitudes of planet signatures observed in our seven new instances are larger than 0.05 dex (up to 0.15 dex), thus favouring the scenario of planet engulfment. An in-depth comparison between our observational abundance results with different models of planet and/or proto-planetary disc formation will provide new, stringent constraints on these proposed scenarios.


\newcommand{\subfigimg}[3][,]{%
  \setbox1=\hbox{\includegraphics[#1]{#3}}
  \leavevmode\rlap{\usebox1}
  \rlap{\hspace*{10pt}\raisebox{\dimexpr\ht1-2\baselineskip}{#2}}
  \phantom{\usebox1}
}

\begin{figure}[hbp]
\centering
\begin{tabular}{c}
    \subfigimg[width=0.75\hsize]{\textbf{a}}{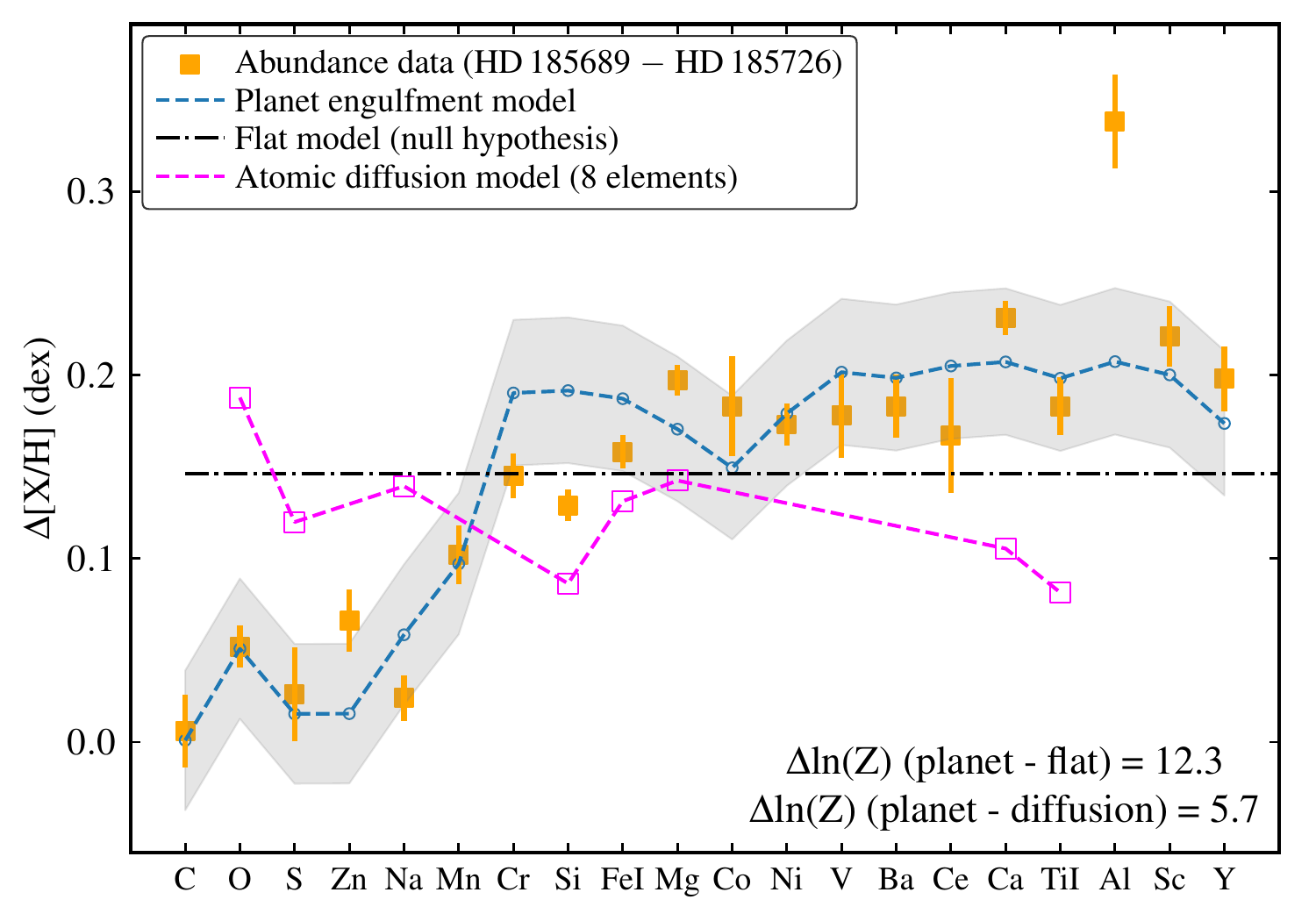} \\
    \subfigimg[width=0.75\hsize]{\textbf{b}}{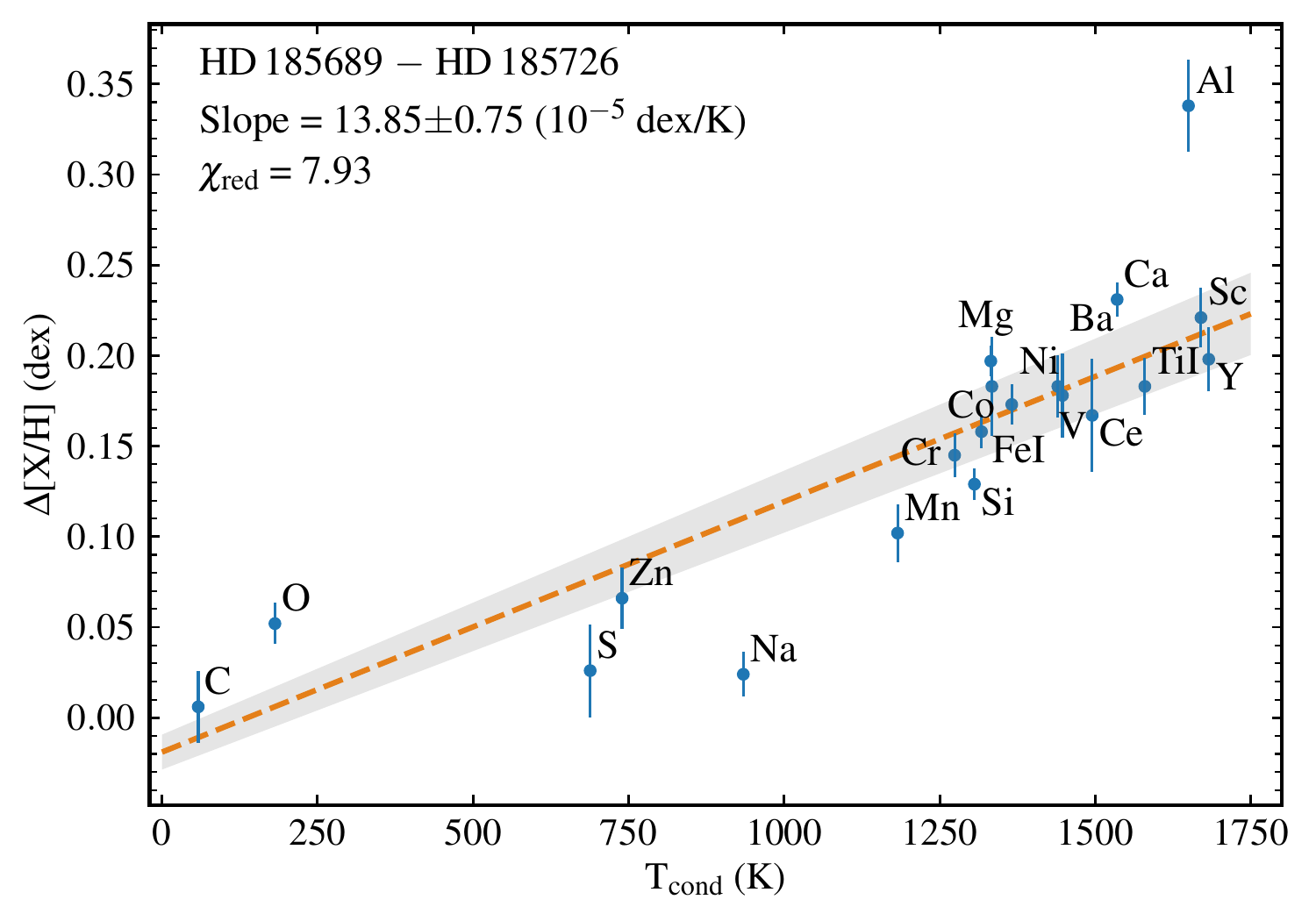} \\
\end{tabular}
\caption{\textbf{Abundance differences ($\Delta$[X/H]) in an example pair HD\,185726/185689 (Pair 124). a.} $\Delta$[X/H] as a function of condensation temperature (\tcond), where the x-axis is labeled with the corresponding elements. The abundance data, best-matched planetary ingestion model with 1\,$\sigma$ posterior probability distribution (shaded region, a term of intrinsic abundance scatter $\sigma_{\rm scatter}$ is included as explained in the Methods), and predicted abundance change due to atomic diffusion for available elements are shown in orange, blue, and magenta, respectively. Bulk Earth material with M$_{\rm E}$ = $3.07^{+0.22}_{-0.26}$ \mearth\ is required to be accreted into HD\,185689's surface, for it to match the observed abundance enhancement relative to that of HD\,185726. \textbf{b.} $\Delta$[X/H] as a function of \tcond. A linear least-squares fit to the abundance data is shown with an orange dashed line. The shaded region represents the 1\,$\sigma$ upper and lower boundary. The error bars are 1\,$\sigma$ uncertainties of the observed abundances. HD\,185689 of this pair exhibits clear chemical signatures of ingestion of planetary material.}
\label{fig:fitting_data}
\end{figure}

\begin{figure}[hbp]
\centering
\begin{tabular}{c}
    \subfigimg[width=\hsize]{\textbf{a}}{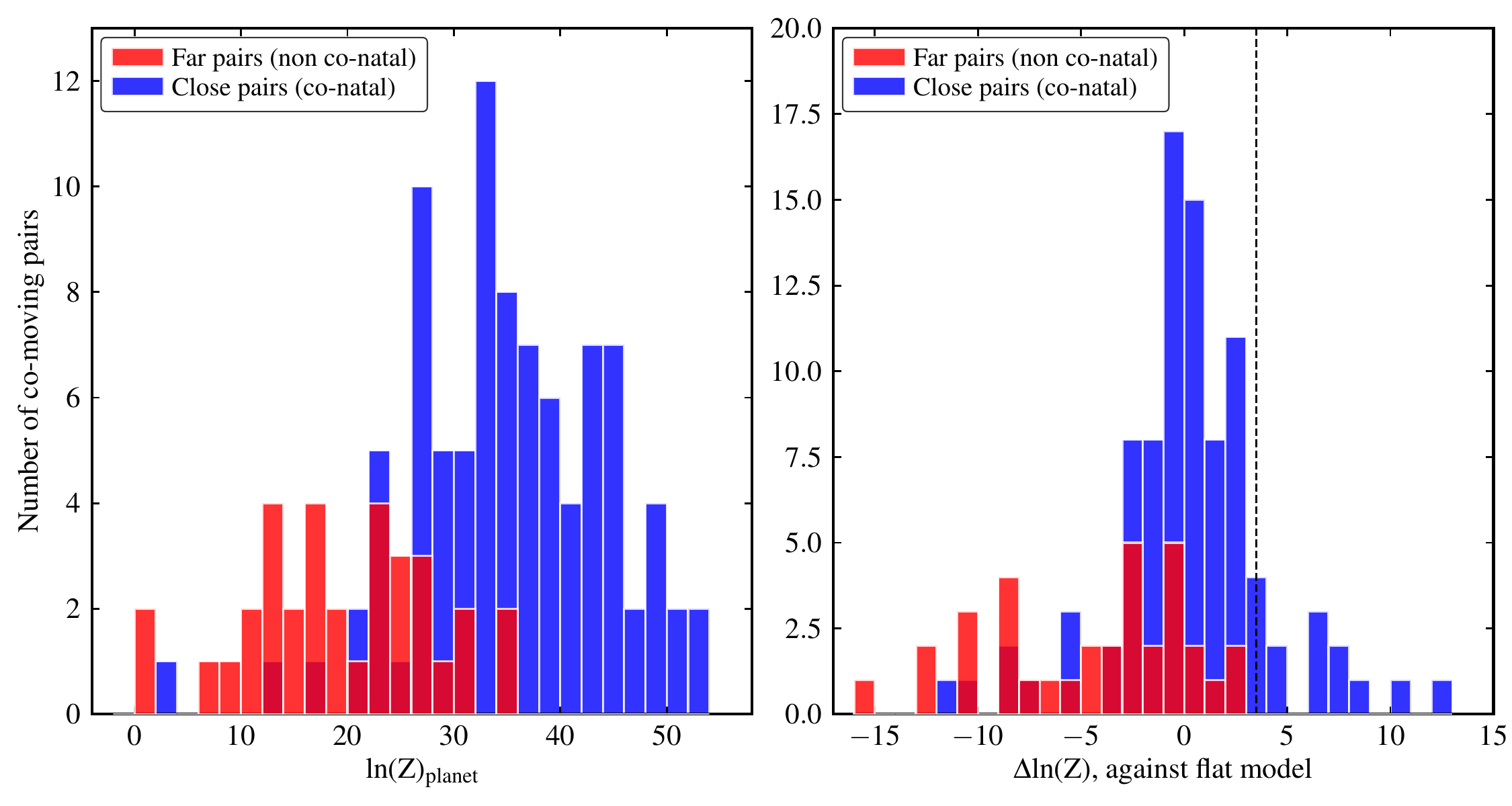} \\
    \subfigimg[width=0.75\hsize]{\textbf{b}}{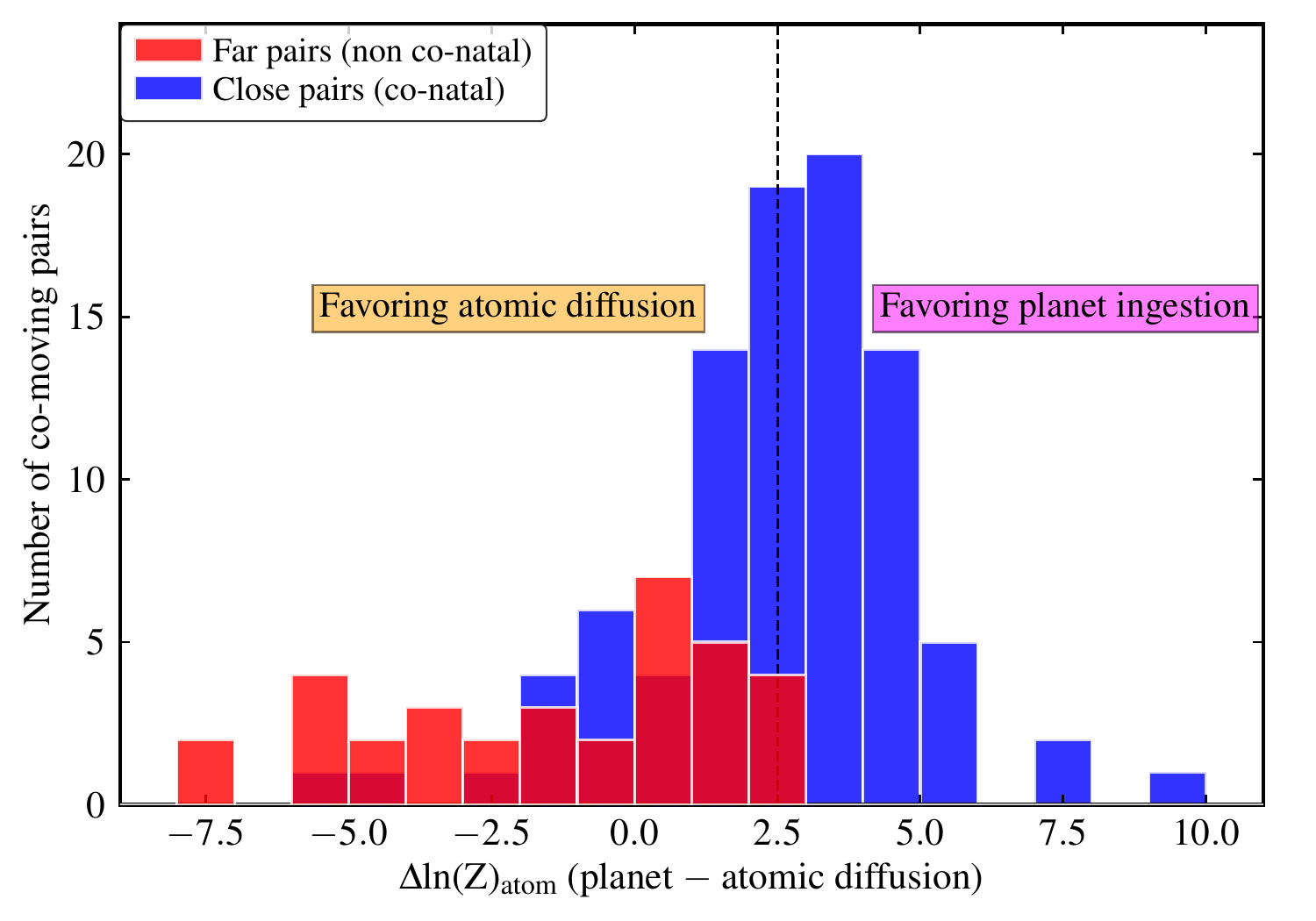} \\
\end{tabular}
\caption{\textbf{The distributions of Bayesian evidence for 125 co-moving pairs. a.} The distributions of Bayesian evidence for planetary ingestion models ln(Z)$_{\rm planet}$ and the differences in Bayesian evidence $\Delta$ln(Z) between the planetary ingestion and flat models. \textbf{b.} The distributions of differences in Bayesian evidence $\Delta$ln(Z)$_{\rm atom}$ between the planetary ingestion and atomic diffusion models for available elements. Red: the control sample of far co-moving pairs (\ds\ $\geq 10^6$\,AU); blue: the target sample of close, co-natal co-moving pairs (\ds\ $< 10^6$\,AU). They show that the co-natal pairs in our sample have more prominent ln(Z)$_{\rm planet}$ and $\Delta$ln(Z), that are distinguishable from the non co-natal pairs.}
\label{fig:hist_dlnZ}
\end{figure}

\begin{figure}[hbp]
\centering
\begin{tabular}{c}
    \subfigimg[width=0.85\hsize]{\textbf{a}}{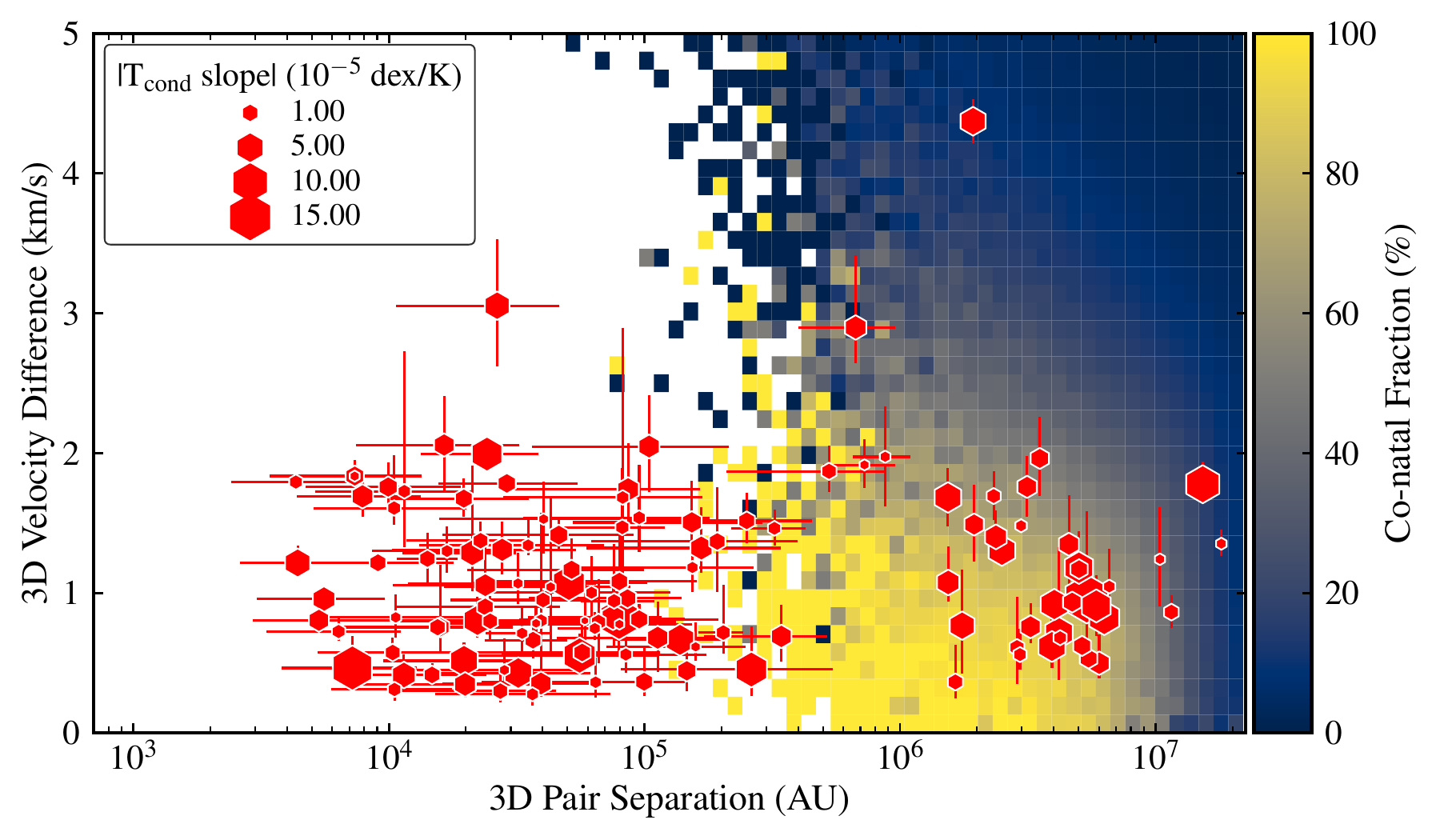} \\
    \subfigimg[width=0.80\hsize]{\textbf{b}}{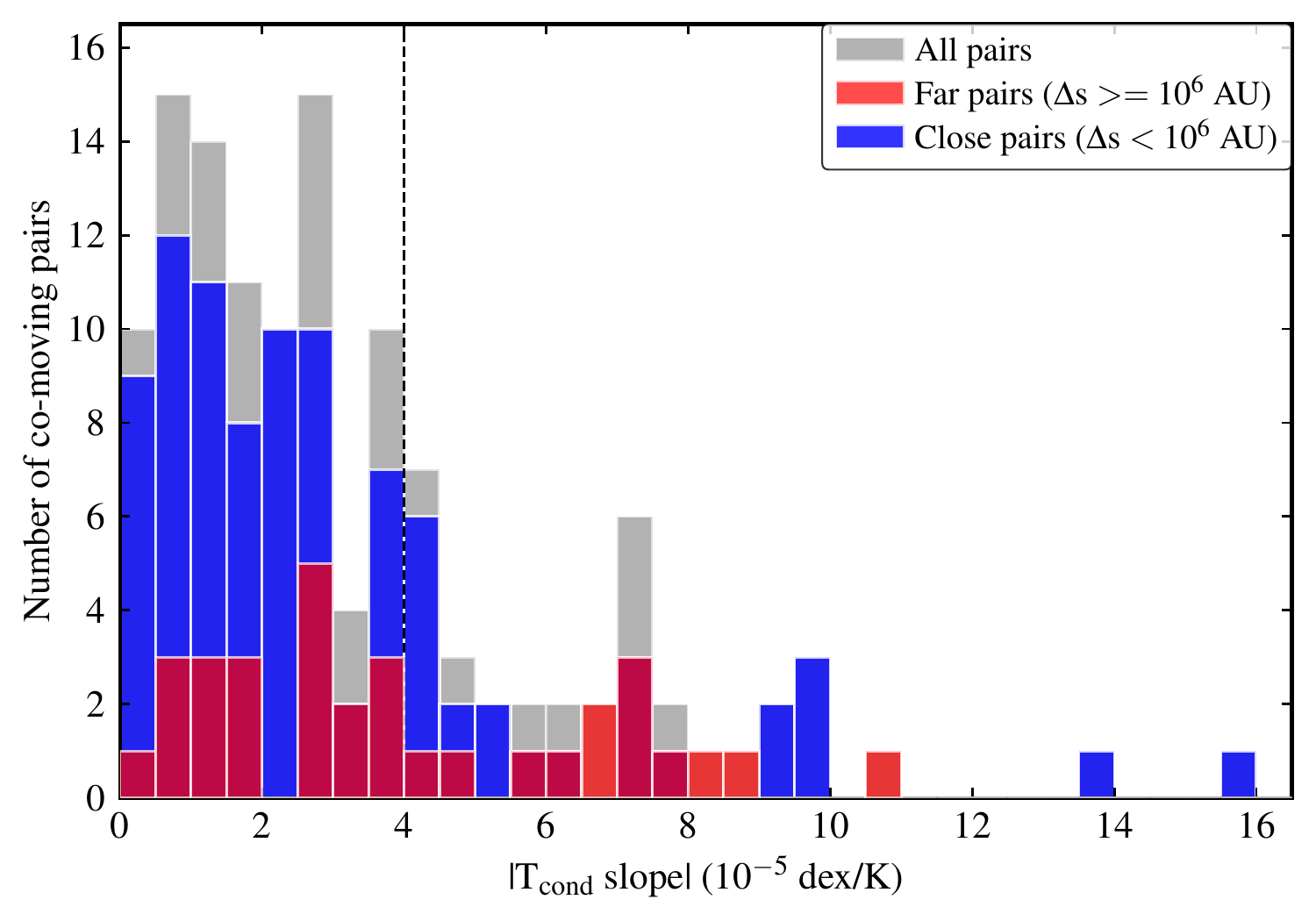} \\
\end{tabular}
\caption{\textbf{The condensation temperature (\tcond) trends for 125 co-moving pairs. a.} The \tcond\ trends in the parameter space of spatial separation (\ds) versus velocity separation (\dv). The sizes of symbols are scaled relative to the absolute value of \tcond\ slopes. The error bars are 1\,$\sigma$ uncertainties in our measurements. Theoretical predictions \citep{kam19} for co-natal fraction of co-moving pairs are shown in the background. The vast majority of pairs with \ds\ between 10$^3$ and 10$^6$ AU are assumed to be co-natal. \textbf{b.} The distributions of absolute slopes of \tcond\ trends. Grey: the whole sample; red: the far co-moving pairs with \ds\ $\geq 10^6$\,AU; blue: the close co-moving pairs with \ds\ $< 10^6$\,AU. They demonstrate that relying solely on the \tcond\ trends cannot efficiently disentangle signatures of planetary ingestion and may lead to false positives, given the distributions are similar between the co-natal pairs and the control sample of far pairs.} 
\label{fig:tcond_slope}
\end{figure}




\newpage
\section*{Methods}

\textbf{Sample selection and data} 

Our sample was selected using photometric and astrometric data from \gaia\ EDR3 \citep{gai21}, while the spatial separations (\ds) and 3D velocity separations (\dv) for the co-moving pairs were calculated and presented in \citep{yon23}. The selection criteria are summarised below: \\
(i) \ds\ $<$ 30\,pc (10$^{6.8}$\,AU) and \dv\ $<$ 2.0\,\kms; \\
(ii) 0.65 $\leq$ (\bprp) $\leq$ 1.15\,mag and $|\Delta$(\bprp)$|$ $\leq$ 0.15\,mag; \\
(iii) $\Delta$M$_G$ $<$ 1\,mag; \\
(iv) \gaia\ G $<$ 10\,mag; \\
(v) A “friends-of-friends” search assuming a connecting threshold of 1\,pc in the 3D distances and omit all groups that have 5 or more members to exclude open clusters and moving groups. 

The details of sample selection, observations, and data reduction were described in \citep{yon23}. In total, we have effective observations of 125 co-moving pairs, i.e., 250 stars. High-resolution, high signal-to-noise ratio (S/N\,$\approx$\,250 per pixel at 600\,nm) spectra of these stars were taken from the following instruments and telescopes: MIKE spectrograph on the Magellan Telescope (7 nights; R $\approx$ 50,000), HIRES on the Keck Telescope (1 night; R $\approx$ 72,000), and UVES on the ESO's Very Large Telescope (26.4 hours; R $\approx$ 110,000). The two stars of each pair in our sample are similar in their colours and absolute magnitudes, facilitating the line-by-line differential approach applied in this study. We note this is the key to achieve high precision in stellar parameters and relative elemental abundances by cancelling major sources of systematic uncertainties from e.g., model atmospheres and atomic line data. Furthermore, the completeness of our sample is $\approx$\,45\% at \gaia\ G $<$ 10\,mag. This represents the largest and most homogeneous sample of co-moving pairs of stars ever examined in a single high-precision spectroscopic analysis. Extended Figure \ref{fig:spectra} shows a portion of reduced spectra of an example pair of stars (HD\,185726/185689; Pair 124) with subtle but clear differences in spectral line strengths of different elements. 

\noindent
\textbf{Stellar parameters} 

The fundamental stellar parameters (i.e., effective temperature \teff, surface gravity \logg, microturbulent velocity \vt, and metallicity [Fe/H]) were derived with high precision \citep{yon23}. Following a similar two-step approach \citep{liu20,liu21}, the absolute stellar parameters were derived relative to the Sun (\teff\ = 5772\,K, \logg\ = 4.44\,cm\,s$^{-2}$, \vt\ = 1.0\,\kms, and [Fe/H] = 0.0 dex) as the first step, and the final differential stellar parameters were derived by comparing the two components of each pair on a line-by-line basis as the second step. 
A grid-search algorithm \citep{liu14} was applied in determining the best solution of stellar parameters for both steps. Excitation and ionisation balance, as well as a zero-difference between \fei\ and \feii\ were imposed in a differential sense for each pair of stars. Our sample includes stars with \teff\ ranging from 5000 to 6500\,K, \logg\ from 3.8 to 4.6, and [Fe/H] from $-$0.6 to 0.4 dex. For differential stellar parameters, the majority of our pairs have $\Delta$\teff $< \pm$ 300\,K, $\Delta$\logg $<$ $\pm$ 0.3\,\,cm\,s$^{-2}$, and $\Delta$[Fe/H] $< \pm$ 0.3 dex. The co-natal sample (91 pairs) and the control sample (34 pairs) cover a similar parameter space, but the differences in stellar parameters within each pair are generally larger in the control sample. For example, the average \teff\ difference is 236\,K for the control sample and 138\,K for the co-natal sample.

The uncertainties in differential stellar parameters are typically 15\,K, 0.035\,\,cm\,s$^{-2}$, and 0.012 dex for \teff, \logg, and [Fe/H], respectively. They do not increase substantially, even for pairs with $\Delta$\teff\ as large as 300\,K, indicating that the systematic errors are effectively suppressed. In addition, 14 common pairs previously analyzed with high-precision spectroscopy \citep{nag20,spi21} agree well in stellar parameters with this study, showing e.g., an average difference of $-$16\,K (standard deviation = 34\,K) for \teff\ and 0.002 dex (standard deviation = 0.009 dex) for [Fe/H], reinforcing their reliability.

\noindent
\textbf{Equivalent width measurements and elemental abundances} 

The equivalent widths (EWs), i.e., strength of spectral lines, of 21 elements (C, O, Na, Mg, Al, Si, S, Ca, Sc, Ti, V, Cr, Mn, Fe, Co, Ni, Zn, Sr, Y, Ba, and Ce) were measured for our sample stars based on a list of carefully selected clean lines \citep{mel12,liu14}, using a semi-automatic code very similar to {\sc REvIEW} \citep{mck22} with each line being visually inspected. We mainly adopted lines with intermediate strengths (10\,m\AA\ $<$ EW $<$ 100\,m\AA), except for a few barium lines, where only strong lines are available. The line-list with atomic data and EW measurements is presented in the Supplementary information. 

The differential abundances for these 21 elements were computed on a line-by-line basis using the established differential stellar parameters. The abundance analysis, like parameter determination, was conducted using the {\sc MOOG2019} software \citep{sne73,sob11} based on 1D local thermodynamic equilibrium (LTE), while the corresponding {\sc ODFNEW} \citep{ck03} stellar model atmospheres were used. Corrections for Hyper-fine structure splitting (HFS) were taken into account for Sc, V, Mn, Co, and Ba, where the atomic data of HFS components were taken from previous studies \citep{kb95,bb15}. 3D non-LTE corrections to the oxygen abundance were applied to the 777\,nm triplet \citep{ama16}. The amount of differential 3D non-LTE abundance corrections for the oxygen triplet can vary between 0.01\,--\,0.08 dex for our co-moving pairs. We also estimated the non-LTE impact on specific elements (Na, Mg, Al, and Mn; \citealp{lin11,ber17,nor17,ber19}) for two pairs (Pair 17 and 22) with $\Delta$\teff\ of $\approx$ 300K and 200K. The effects ($\lesssim$ 0.01\,--\,0.015 dex) are minor due to our differential approach and do not affect the majority of the co-natal sample. 

The differences in elemental abundances ($\Delta$[X/H]) for the 125 co-moving pairs are presented in the Supplementary information. The uncertainties in differential abundances ($\sigma_{\rm \Delta[X/H]}$) were computed following the algorithm presented in previous studies \citep{ben14,liu20}. They were calculated by adding in quadrature the errors introduced from the uncertainties of stellar atmospheric parameters, and the standard errors of the line-to-line abundance scatter. The errors in the differential elemental abundances are below 0.02 dex for most species for most of our sample pairs. We note the average errors in differential abundances for almost all our co-moving pairs are only $\sim$ 0.015 dex (3.5\%), a substantial improvement over the traditional abundance analysis with a typical error of 0.05\,--\,0.1 dex (12\,--\,26\%). We emphasize that the small uncertainties are due to: (i) the high resolution and high S/N spectra; (ii) the line-by-line differential analysis that suppress significantly the systematic uncertainties, and (iii) the sample selection which ensured that for a given pair of stars, the stellar parameters are similar (e.g., $\Delta$\teff\ within 200\,K for most of the co-natal pairs). 

\noindent
\textbf{Validation of abundance results: internal and external comparison} 

We have multiple observations of seven co-moving pairs (14 stars). Six pairs were observed twice on consecutive nights with the Magellan Telescope, and one pair (Pair 38) was observed with both the Magellan Telescope and the Keck Telescope. Although the final results of these pairs (both stellar parameters and elemental abundances) were based on the combined observations, we also analysed each set of observations independently. This enables us to test and compare our internal abundance results quantitatively. As shown in Extended Figure \ref{fig:comparison}\,\textbf{a}, the deviations in differential abundances ($\Delta$[X/H]) between multiple observations for these 7 pairs are centered around zero with a standard deviation of $\approx$\,0.02 dex, which agree with the corresponding average abundance uncertainties. The agreement suggests that our differential approach has effectively suppressed systematics from different observing conditions and instruments. 

In addition, a common pair HD\,133131A/B was observed with Magellan/MIKE in this study, and with VLT/UVES \citep{liu21}. Both sets of data were analysed independently, but using a very similar method and line-list. This common pair thus enables us to provide a further, external test in validating our results of differential abundances. The comparison of differential abundances for this pair is shown in Extended Figure \ref{fig:comparison}\,\textbf{b}. The average abundance difference between this study and \citep{liu21} is zero within uncertainties ($<$$\Delta$[X/H]$>$ = 0.003\,$\pm$\,0.004 dex, with the standard deviation = 0.019 dex), indicating that systematics from different instruments and telescopes were significantly suppressed and have essentially no influence on our differential abundance results, even at the precision level of 0.01\,--\,0.02 dex (2--5\%). 

\noindent
\textbf{Description of Bayesian method} 

We developed an independent test for examining the observed abundance differences within a given pair of stars and their potential origin via Bayesian analysis. 
In general, for a model defined by a set of parameters ($\theta$) in explaining the observing results ($D$), we can efficiently sample from the posterior probability distributions of those parameters, find the parameter estimation, and calculate the joint posterior on the whole set of parameters from Bayes theorem: \\
\begin{equation}
P(\theta|D) = \frac{P(D|\theta)P(\theta)}{P(D)}
\end{equation}
where P($D|\theta$) is the likelihood, P($\theta$) is the prior, P($D$) is the Bayesian evidence, i.e., the integral of the posterior over the parameter, and P($\theta|D$) is the posterior. 
A nested sampling algorithm was applied to estimate simultaneously the Bayesian evidence and the posterior for given models using {\sc dynesty} \citep{spe20}. 

For the simplified model of planetary ingestion, we examined the mass of bulk Earth composition (M$_{\rm E}$) required to be ingested into the stellar convection zone (\mcz) of one of the stars to match the abundance pattern for a given co-moving pair. The model for Bayesian parameter fitting is described as below: \\
\begin{equation}
M_{X, \rm CZ} = \frac{10^{\rm [X/H]}m_X}{\sum_X 10^{\rm [X/H]}m_X} f_{\rm CZ} M_{\rm star}
\end{equation}
\begin{equation}
M_{X, \rm E} = f_{X, \rm E} M_{\rm E}
\end{equation}
\begin{equation}
\Delta[X/H] = log_{10} \frac{M_{X, \rm E} + M_{X, \rm CZ}}{M_{X, \rm CZ}}
\end{equation}
where M$_{X, \rm CZ}$ is the mass of element X in the stellar convection zone; m$_X$ is the mass number of element X; $f_{\rm CZ}$ is the mass fraction of convection zone, $f_{X, \rm E}$ is the mass fraction for element X being accreted assuming the bulk Earth composition \citep{asp09}, M$_{X, \rm E}$ is the total mass of element X introduced by planetary ingestion. We note the absolute metallicity of each co-moving pair was taken from the star with lower [Fe/H]. The mass range of each star was estimated based on its stellar parameters. 
The final adopted values of $f_{\rm CZ}$ were estimated by interpolation between a grid of $f_{\rm CZ}$ with corresponding \teff, which was calculated using the {\sc MESA} program (see details in \citealp{yon23}). We find the correlation between $f_{\rm CZ}$ and \teff\ from {\sc MESA} is similar to that as presented in previous studies \citep{pin01}. Additionally, changing $f_{\rm CZ}$ mainly affects the inferred mass of planetary material (M$_{\rm E}$) being accreted, but not the relative Bayesian evidence that we adopted to identify the candidates. 

For each pair, the planetary ingestion model with a set of parameters $\theta$ $\sim$ \{M$_{\rm E}$, $\sigma_{\rm scatter}$\} ($\sigma_{\rm scatter}$: intrinsic abundance scatter) was generated to fit the abundance data $D$ $\sim$ \{$\Delta$[X/H], $\sigma_{\rm \Delta[X/H]}$\} ($\sigma_{\rm \Delta[X/H]}$: measured uncertainties). $\sigma_{\rm scatter}$ was introduced to avoid overfitting and to account for additional systematic errors. Uniform probability distributions with proper boundaries were defined for the prior on these parameters: M$_{\rm E}$ $\sim$ (0.5 -- 80 \mearth); $\sigma_{\rm scatter}$ $\sim$ (0 -- 1), same for all elements. A standard likelihood function for a set of observation $i$ was given by a Gaussian distribution of the form below (in logarithmic scale): \\ 
\begin{equation}
{\rm log} L = norm - {\sum_{j} ^N} \frac{(\Delta\rm{[X_j/H]_{\rm observation,i}} - \Delta\rm{[X_j/H]_{\rm model,i}})^2}{2 (\sigma_{\rm \Delta[X_j/H],\mathit i}^2 + \sigma_{\rm scatter,\mathit i}^2)}
\end{equation}
where $N$ is the number of elements; the logarithmic normalisation term $norm$ is defined as: \\
\begin{equation}
norm = - \frac{{\rm log} (2 \pi)}{2} - {\sum_{j} ^N} {\rm log} \sqrt{(\sigma_{\rm \Delta[X_j/H],\mathit i}^2 + \sigma_{\rm scatter,\mathit i}^2)}
\end{equation}
After setting up the prior distributions and likelihood function, we then run the sampler using 256 live points with a stopping criterion of 0.05 for each co-moving pair. The final value of Bayesian evidence ln(Z)$_{\rm planet}$ (marginal likelihood) and its uncertainties, as well as the posterior samples of fitting parameters can be calculated. The posterior distribution in our case is predominantly unimodal, ensuring robust convergence of the Bayesian evidence. Random bootstrap sampling reinforces this assertion, inducing a mere 5\% variation in the evidence (within detection limits). In addition, a combination of carbonaceous chondrite and bulk Earth composition were tested but no clear differences in the Bayesian evidence were found. 

Similarly, for the flat model (null hypothesis), we fitted our abundance data to a base model with a set of parameters $\theta$ $\sim$ \{$\Delta$[M/H]$_{\rm offset}$, $\sigma_{\rm scatter}$\} ($\Delta$[M/H]$_{\rm offset}$: overall abundance offset). The prior distributions are defined as uniform with: $\Delta$[M/H]$_{\rm offset}$ $\sim$ ($-$0.5 -- 0.5 dex); $\sigma_{\rm scatter}$ $\sim$ (0 -- 1). The form of the likelihood function and the other sampler settings are the same as those for the planetary ingestion model, except here we set $\Delta$[X$_j$/H]$_{\rm model, \mathit i}$ = $\Delta$[M/H]$_{\rm offset, \mathit i}$ for all elemental abundances. The Bayesian evidence ln(Z)$_{\rm flat}$ and the posterior of the corresponding fitting parameters for each co-moving pair were calculated. The difference in Bayesian evidence $\Delta$ln(Z) between the planetary ingestion and the flat model provides us with an independent indicator to effectively distinguish the potential chemical signatures of ingestion of planetary material. 

We then generated a mock ``signal" data sample of 125 pairs of stars, where the abundance differences within each pair of stars (with inherited measurement uncertainties) were simulated based on the predicted parameters $\theta$ drawn from the best-fitting model, representing the case of noise-free detection. Furthermore, a mock ``noise" data sample with zero abundance difference and random noise drawn from the corresponding abundance uncertainties ($\sigma_{\rm \Delta[X/H]}$) were generated for the 125 co-moving pairs. Bayesian analyses with the same approach as above were applied to the mock signal and noise data sample. The distributions of $\Delta$ln(Z) between planetary ingestion and flat model for our sample (far and close pairs), mock signal, and mock noise data sample are shown in Extended Figure \ref{fig:hist_dlnZ_all}. It demonstrates that the Bayesian modelling in our study can efficiently distinguish potential planet signatures from the realistic noise and pure signal (especially for the 91 co-natal pairs), when a cutoff value of 3.5 for $\Delta$ln(Z) is adopted. 

\noindent
\textbf{Effects of atomic diffusion} 

Stellar atomic diffusion combines two opposing processes: gravitational settling and radiative acceleration \citep{mic84}, which can alter a star's surface abundance at different evolutionary stages \citep{dot17}. Recent studies have revealed subtle effects of atomic diffusion in both open clusters \citep{liu19} and wide binaries \citep{liu21}. In principle, the effects of atomic diffusion are more likely imprinted into stars with relatively large $\Delta$\teff\ and/or $\Delta$\logg. 

In this study, we applied a Bayesian analysis with a similar approach as described above to examine the effects of atomic diffusion in our sample. The atomic diffusion models \citep{dot17} for 8 available elements (O, Na, Mg, Si, S, Fe, Mg, and Ti) were taken into account (O and S were not measurable for a few pairs). Firstly, the predicted abundance change/difference ($\Delta$[X/H]$_{\rm atom}$) between two stars of a given pair can be determined via interpolation of the model grid with established atmospheric parameters and estimated stellar mass range. The two stars of most pairs have similar mass (difference $<$\,10\%). Secondly, we fitted our abundance data to a base model (similar to the flat model) with a set of parameters $\theta$ $\sim$ \{$\Delta$[M/H]$_{\rm offset}$, $\sigma_{\rm scatter}$\}. The prior distributions, the form of likelihood function, and the other sampler settings remain consistent with those of the flat model (but applied to 6--8 available elements), except for $\Delta$[X$_j$/H]$_{\rm model, \mathit i}$ = $\Delta$[X$_j$/H]$_{\rm atom, \mathit i}$ + $\Delta$[M/H]$_{\rm offset, \mathit i}$. Thirdly, the Bayesian evidence for the atomic diffusion model ln(Z)$_{\rm atom}$ for each co-moving pair was derived and compared to that for the planetary ingestion model for those 6--8 available elements only. The differences in Bayesian evidence between the planetary ingestion and atomic diffusion model $\Delta$ln(Z)$_{\rm atom}$ were calculated, in order to distinguish the effects of atomic diffusion from potential planet signatures. The distributions of $\Delta$ln(Z)$_{\rm atom}$ for our sample, as well as $\Delta$ln(Z)$_{\rm atom}$ as a function of spatial separations (\ds), are shown in Figure \ref{fig:hist_dlnZ}\,\textbf{b} and Extended Figure \ref{fig:dlnZ_atom_ds}, demonstrating that $\Delta$ln(Z)$_{\rm atom}$ as an indicator, unlike that of \tcond\ trends, distinguishes our co-natal sample from the control sample of co-moving pairs. The results also enable us to tentatively define a further criterion of $\Delta$ln(Z)$_{\rm atom}$ $>$ 2.5 to identify candidates of planetary ingestion which are not likely affected by atomic diffusion. Our results are unlikely to change substantially when using different atomic models because the recipes and formula are similar, especially for relative abundance differences where the zero-point is less significant. 

\noindent
\textbf{Relative abundances and condensation temperature} 

We examined the correlation between relative elemental abundances and their \tcond, i.e., \tcond\ trend for each sample pair with different fitting ranges. The co-moving pairs of stars have an average $|$\tcond\ slope$|$ = $(3.5 \pm 0.3) \times 10^{-5}$ dex\,K$^{-1}$ when fitting all elements. Similarly, the linear least squares fit was also applied to refractory elements only (\tcond\ $>$ 1000\,K), and to all elements excluding C and O because we are aware that the \tcond\ trends are especially sensitive to those elements. Although the results for individual pairs may vary, the distributions of \tcond\ slopes are essentially unchanged when employing different fitting ranges of \tcond. For our candidates with large $\Delta$ln(Z) ($>$\,3.5), they also exhibit significant \tcond\ trends with an average $|$\tcond\ slope$|$ = $(6.9 \pm 1.0) \times 10^{-5}$ dex\,K$^{-1}$, which is larger than the pairs without detection in general. However, as shown in Figure \ref{fig:tcond_slope}, co-moving pairs of stars with large \tcond\ trends (e.g., in the control sample) can be false positives due to random abundance variation and/or atomic diffusion. Therefore the relative Bayesian evidence $\Delta$ln(Z) is primarily adopted in this study. 


\noindent
\textbf{Ingestion of Earth-like material and the inferred mass} 

The total mass of accreted Earth material can be estimated from the best-fitting planetary ingestion model. We plotted the accreted mass (in units of \mearth) as a function of $\Delta$ln(Z) in Extended Figure \ref{fig:mearth_dlnZ}. As expected, the mass of accreted Earth material increases qualitatively with increasing $\Delta$ln(Z). We note it is not a linear correlation because the mass of stellar convective envelope \mcz\ differs between stars with different \teff\ and mass range. For hotter stars with thinner convective envelope, the required amount of ingestion of planetary material to match the observing abundance differences would be smaller than the stars with similar abundance differences but larger convection zone. The mass fraction of convection zone of a star (f$_{\rm CZ}$) thus plays an important role in determining the factor of surface abundance change due to potential planetary ingestion. We also examined the mass of accreted Earth material for the seven candidates and found the average to be 4.3 $\pm$ 0.8 \mearth\ (with the standard deviation of 2.1 \mearth). The final fitting results of these co-moving pairs are presented in Extended Table \ref{table:fitting_results}. 


\noindent
\textbf{Timescales of planet formation and engulfment} 

Previous studies \citep{mel09,cha10} proposed that terrestrial planet formation in our Solar system removed refractory material from the proto-planetary disc, which was then accreted by the Sun. In this scenario, a deficiency in refractory elements was expected in stars with rocky planets compared to their stellar twins (i.e., stars with identical atmospheric parameters) without planets. However, rocky planet formation ($<$\,10 Myr) occurs when the host star is mostly convective, making it difficult to imprint detectable stellar chemical signatures. 

Alternatively, planet engulfment can enhance the stellar surface composition in a way that mirrors the chemical composition of the ingested material \citep{pin01}, leading to the trend between abundance differences and condensation temperature. N-body simulations show that collisions and instabilities in stellar-planetary systems are more frequent within the first 100 Myr \citep{izi21,bi23}, a timescale comparable to when the star reaches the main-sequence and reduces its convection zone to the current size, suggesting that planet engulfment might imprint a detectable signature at this stage. The strength and duration of the signature can be influenced by physical processes in the host star, such as thermohaline mixing \citep{tv12}, which is highly uncertain. Additionally, later accretion events (e.g., due to outer perturbers or the disruption of inner super-Earths' atmospheres) can occur on timescales of 100 Myr to 1 Gyr \citep{mo20}, capable of imprinting distinct stellar chemical signatures.

\backmatter

\bmhead{Data availability}
The spectral data underlying this article are available in Keck Observatory Archive (\url{https://koa.ipac.caltech.edu/cgi-bin/KOA/nph-KOAlogin}) and ESO Science Archive Facility (\url{http://archive.eso.org/eso/eso_archive_main.html}). They can be accessed with Keck Program ID: W244Hr (Semester: 2021B, PI: Liu) and ESO Programme ID: 108.22EC.001 (PI: Yong), respectively. The spectra from the Magellan Telescope can be shared upon request to the corresponding author. The rest of data underlying this article are available in the article and in its online supplementary material.

\bmhead{Code availability}
The stellar line analysis program MOOG is available at \url{https://www.as.utexas.edu/~chris/moog.html}. The stellar model atmospheres are available at \url{http://kurucz.harvard.edu/grids.html}. The code for equivalent width measurements is very similar to {\sc REvIEW}, which is provided in \url{https://github.com/madeleine-mckenzie/REvIEW}. The Bayesian modelling program {\sc dynesty} is available at \url{https://github.com/joshspeagle/dynesty}.

\bmhead{Acknowledgements}
This research were partly supported by the Australian Research Council Centre of Excellence for All Sky Astrophysics in 3 Dimensions (ASTRO 3D), through project number CE170100013. Y.S.T.\ acknowledges financial support from the Australian Research Council through DECRA Fellowship DE220101520. M.T.M.\ acknowledges the support of the Australian Research Council through Future Fellowship grant FT180100194. B.B.\ thanks the European Research Council (ERC Starting Grant 757448-PAMDORA) for their financial support. M.J.\ gratefully acknowledges funding of MATISSE: \textit{Measuring Ages Through Isochrones, Seismology, and Stellar Evolution}, awarded through the European Commission's Widening Fellowship. This project has received funding from the European Union's Horizon 2020 research and innovation programme. We thank Dr.~Alex Ji for offering valuable advice on data collection and preparation. We thank Dr.~Simon Campbell, Dr.~Alex Mustill, and Dr.~Qinghui Sun for their insightful discussions. We thank the referees for their careful reading and constructive comments on this manuscript. 
The C3PO program is made possible through the Carnegie Observatories' support and allocation of observation time on the Magellan telescope. The authors wish to recognize and acknowledge the very significant cultural role and reverence that the summit of Maunakea has always had within the indigenous Hawaiian community. We are most fortunate to have the opportunity to conduct observations from this mountain. This work has made use of data from the European Space Agency (ESA) mission Gaia (\url{https://www.cosmos.esa.int/gaia}), processed by the Gaia Data Processing and Analysis Consortium (DPAC, \url{https://www.cosmos.esa.int/web/gaia/dpac/consortium}). Funding for the DPAC has been provided by national institutions, in particular the institutions participating in the Gaia Multilateral Agreement. 

\bmhead{Author contributions}
F.L. led and played a part in all aspects of the observations and data analysis for this study, wrote and developed the manuscript. Y.S.T. initiated the C3PO program, carried out the Magellan observations, and contributed to the statistical analysis of the research. D.Y. carried out the observations and part of spectroscopic analysis of the Magellan and the VLT data, and contributed in designing this study. B.B., M.J., and A.D. contributed in the theoretical interpretations of the observational results. A.K., M.M., and F.D. contributed in the development and writing of the paper. All authors read, commented, and agreed on the manuscript.

\bmhead{Competing interests}
The authors declare no competing interests.

\noindent\textbf{Additional information}

\noindent\textbf{Correspondence and requests for materials} should be addressed to F. Liu.

\noindent\textbf{Peer review information}

\noindent\textbf{Reprints and permissions information} 

\smallskip
\noindent\textbf{Supplementary information}

The spectral line-list used in this study, along with atomic data and equivalent width (EW) measurements, is available for review in its entirety as an ASCII file (``ews\_all.csv"). 

The adopted elemental abundance differences ($\Delta$[X/H]) and the associated uncertainties for each co-moving pair of stars in this study are presented in its entirety as an ASCII file (``abun\_all.csv"). 



\newpage
\section*{Extended Data}

\setcounter{figure}{0}
\renewcommand{\figurename}{Extended Fig.}

\begin{figure}[hbp]
\includegraphics[width=\hsize]{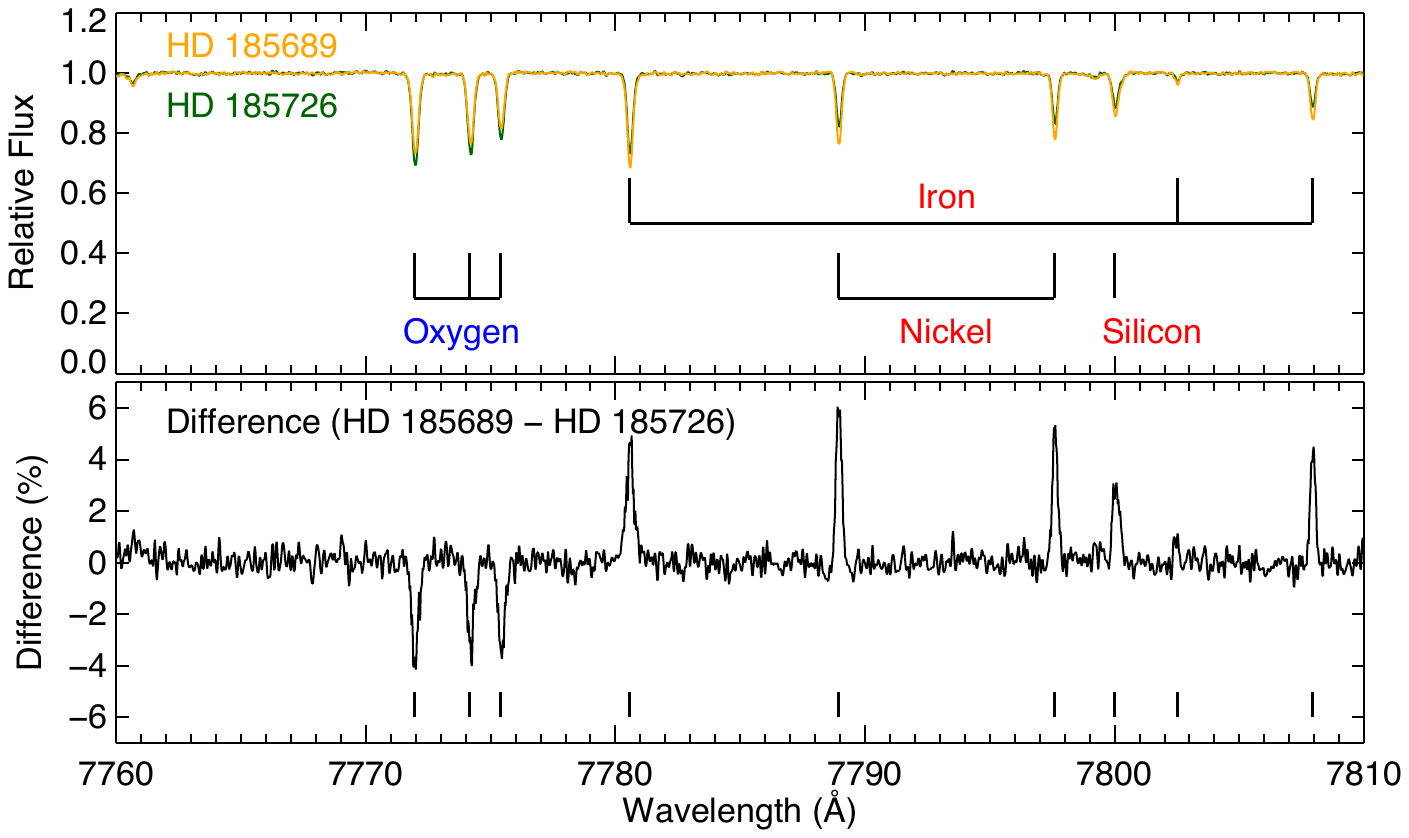}
\caption{\textbf{A portion of the reduced spectra of an example co-moving pair of stars (HD\,185726/185689; Pair 124).} The spectra of HD\,185726 and HD\,185689 are plotted in dark green and orange, respectively. The stellar parameters of these two stars are: [\teff\ = 6271\,K; \logg\ = 4.2\,cm\,s$^{-2}$; [Fe/H] = $-$0.364\,dex] for HD\,185726, and [\teff\ = 6132\,K; \logg\ = 4.36\,cm\,s$^{-2}$; [Fe/H] = $-$0.207\,dex] for HD\,185689. Representative lines of oxygen, iron, silicon, and nickel adopted in this analysis are marked out. It demonstrates that even at the level of a few percent, the differences in line strengths (for different elements) between two stars can be clearly revealed, if they exist.}
\label{fig:spectra}
\end{figure}

\begin{figure}[hbp]
\centering
\includegraphics[width=\columnwidth]{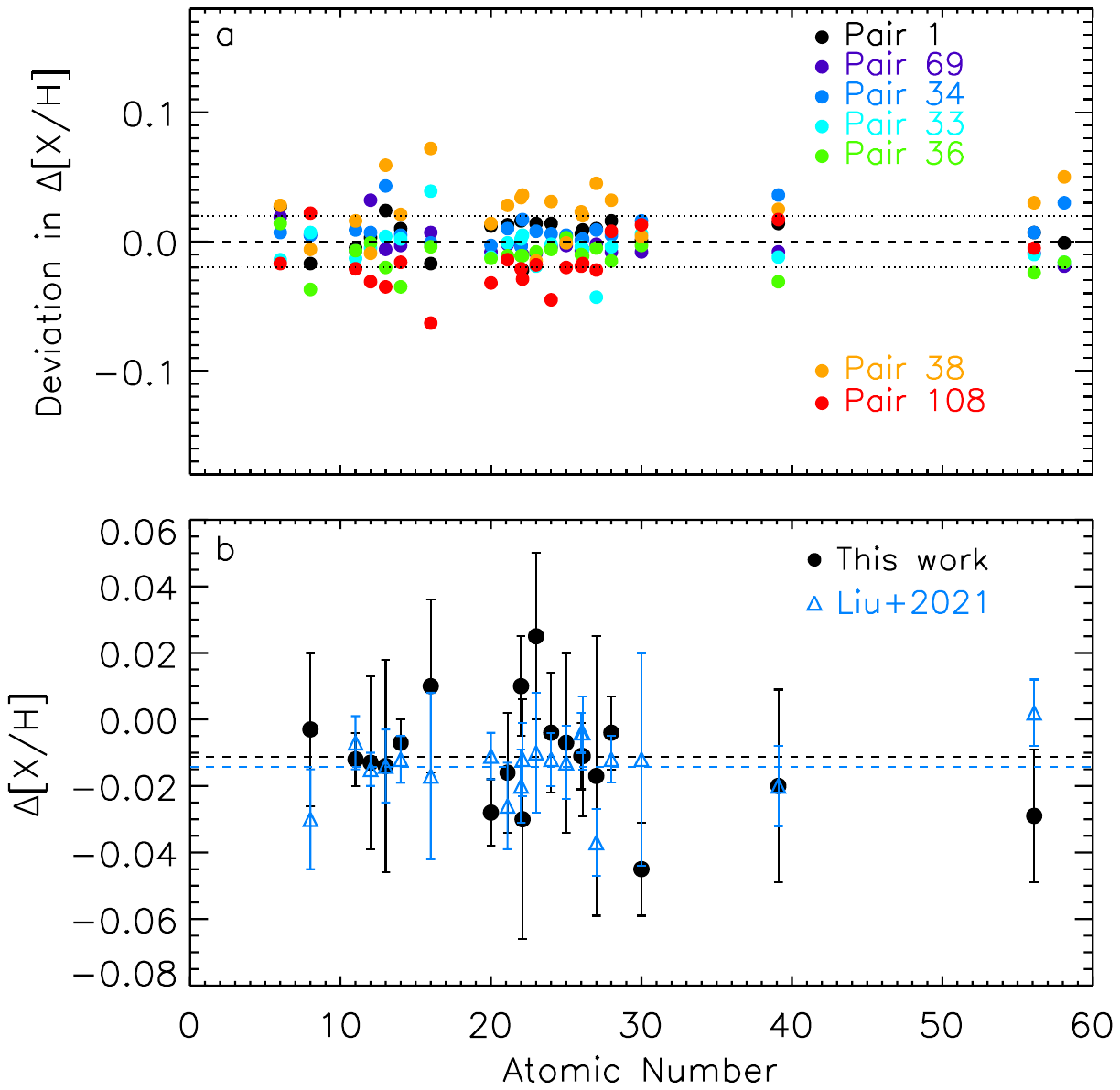}
\caption{\textbf{Comparison of abundance results. a.} The deviations in differential abundances ($\Delta$[X/H]) as a function of atomic number for seven pairs with multiple observations. Different colours represent the deviations in $\Delta$[X/H] for different pairs, as specified in the legend. The dashed lines mark out the 1\,$\sigma$ range around zero. \textbf{b.} $\Delta$[X/H] as a function of atomic number for a common pair HD\,133131A/B between this study and \citep{liu21}. Black circles and blue rectangles represent the abundance results from two independent observations. They demonstrate that the pairwise differences are nearly zero with the standard deviation of $\approx$\,0.02 dex. We note that two pairs (38 and 108) exhibit slightly larger differences (still within 0.03\,--\,0.04 dex for most elements) between the two observations, possibly due to different instruments (for Pair 38) and varying S/N achieved (100--150 versus 200 for Pair 108).}
\label{fig:comparison}
\end{figure}

\begin{figure}[hbp]
\centering
\includegraphics[width=\columnwidth]{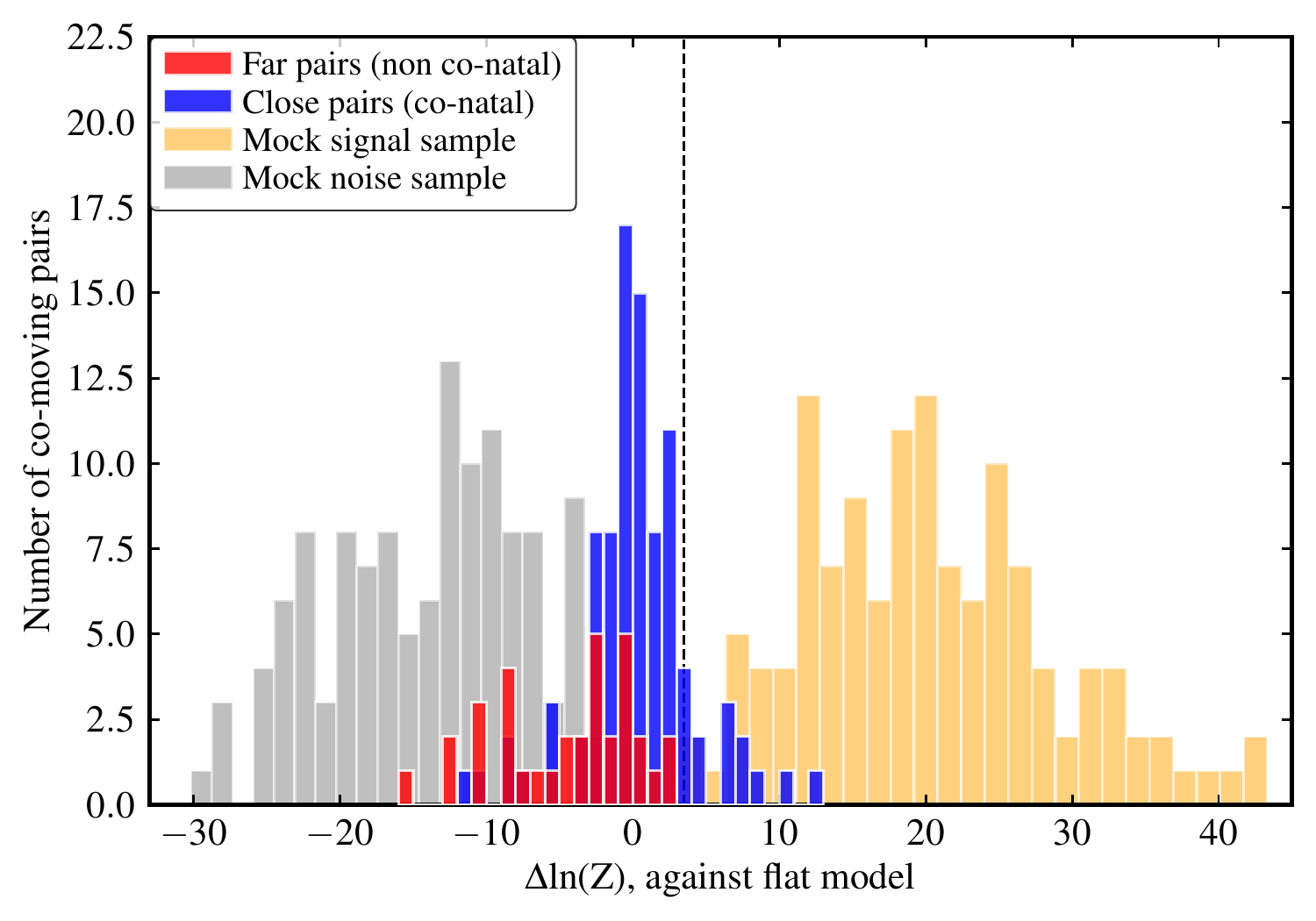}
\caption{\textbf{The distributions of differences in Bayesian evidence $\Delta$ln(Z) between the planetary ingestion and the flat models.} Red: the control sample of far co-moving pairs (\ds\ $\geq 10^6$\,AU); blue: the target sample of close, co-natal co-moving pairs (\ds\ $< 10^6$\,AU); grey: the mock noise sample; orange: the mock signal sample. The distributions, unlike that of \tcond\ trends, are distinguishable between different samples, indicating that we can effectively disentangle the potential planet signatures in the 91 co-natal pairs from the control sample of far pairs, as well as from the mock sample representing realistic noise and pure signal.}
\label{fig:hist_dlnZ_all}
\end{figure}

\begin{figure}[hbp]
\centering
\includegraphics[width=0.95\hsize]{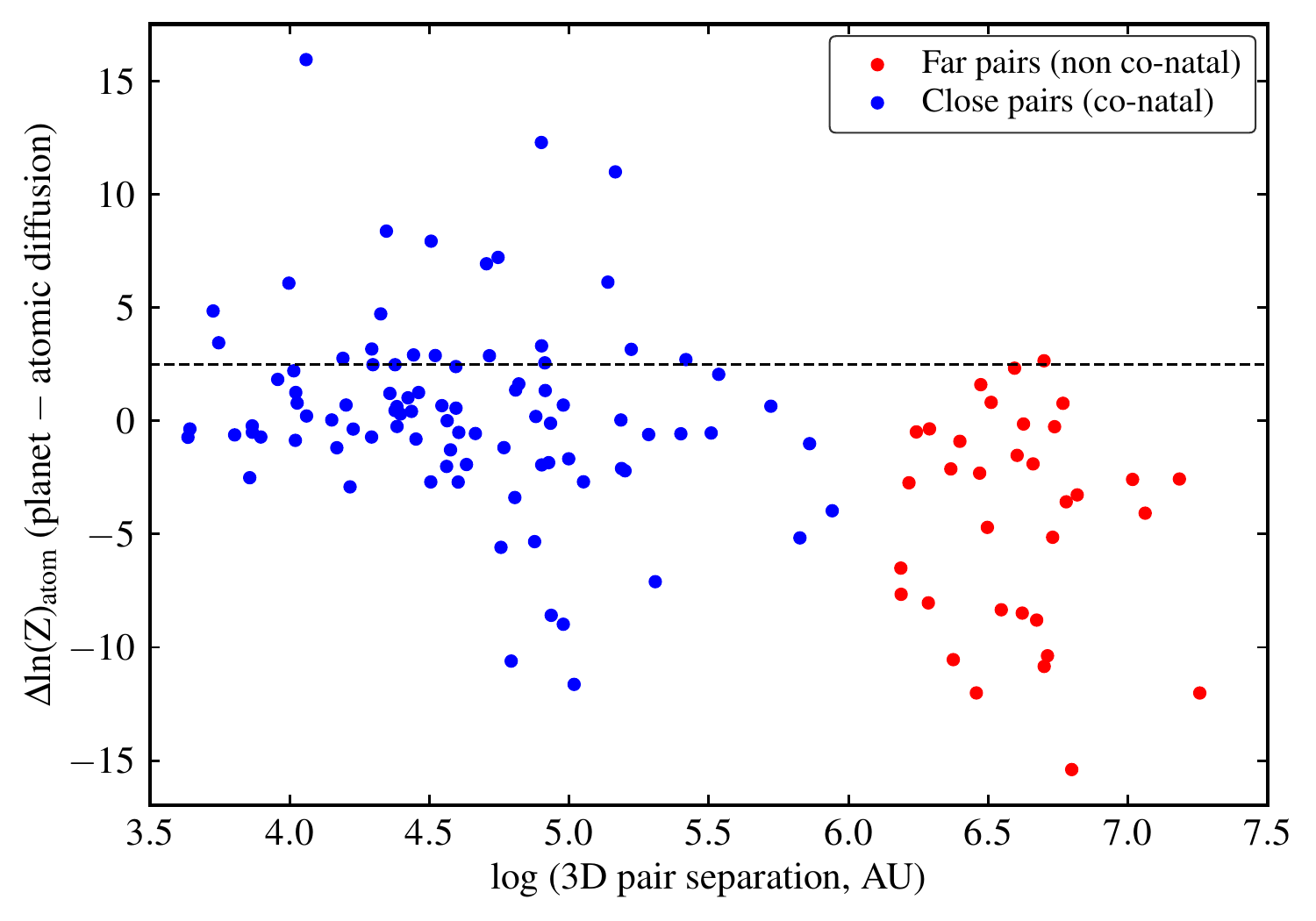}
\caption{\textbf{Differences in Bayesian evidence $\Delta$ln(Z)$_{\rm atom}$ between the planetary ingestion and the atomic diffusion models as a function of spatial separations \ds}. The red and blue circles represent the far and close co-moving pairs, respectively. The dashed line marks out our selection criterion for $\Delta$ln(Z)$_{\rm atom}$. It demonstrates that the far co-moving pairs (non co-natal sample) are distinctively affected by atomic diffusion (possibly due to larger differences in the relative stellar parameters such as $\Delta$\teff), when compared to the close, co-natal co-moving pairs.}
\label{fig:dlnZ_atom_ds}
\end{figure}

\begin{figure}[hbp]
\centering
\includegraphics[width=\columnwidth]{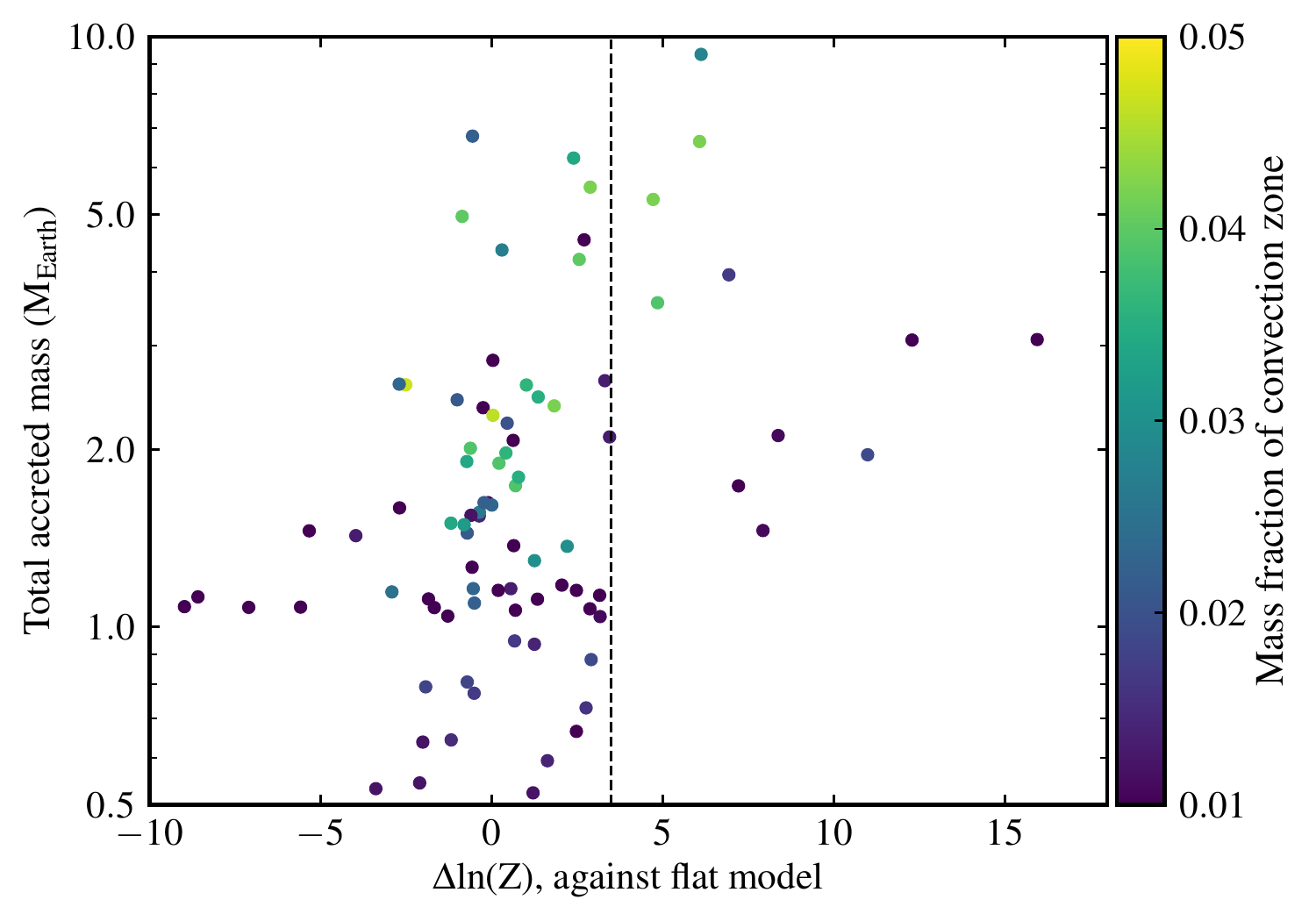}
\caption{\textbf{Total accreted mass of Earth material (from the best-fitting model) as a function of $\Delta$ln(Z) between the planetary ingestion and the flat models for the 91 close, co-natal co-moving pairs.} The dashed line marks out our selection criterion for $\Delta$ln(Z). The data are colour-coded with the mass fraction of convection zone ($f_{\rm CZ}$), indicating that stars with larger $\Delta$ln(Z) tend to have larger accreted mass, while the exact amount of accretion is determined by $f_{\rm CZ}$.}
\label{fig:mearth_dlnZ}
\end{figure}

\begin{figure}[hbp]
\centering
\begin{tabular}{c}
    \subfigimg[width=0.9\hsize]{\textbf{a}}{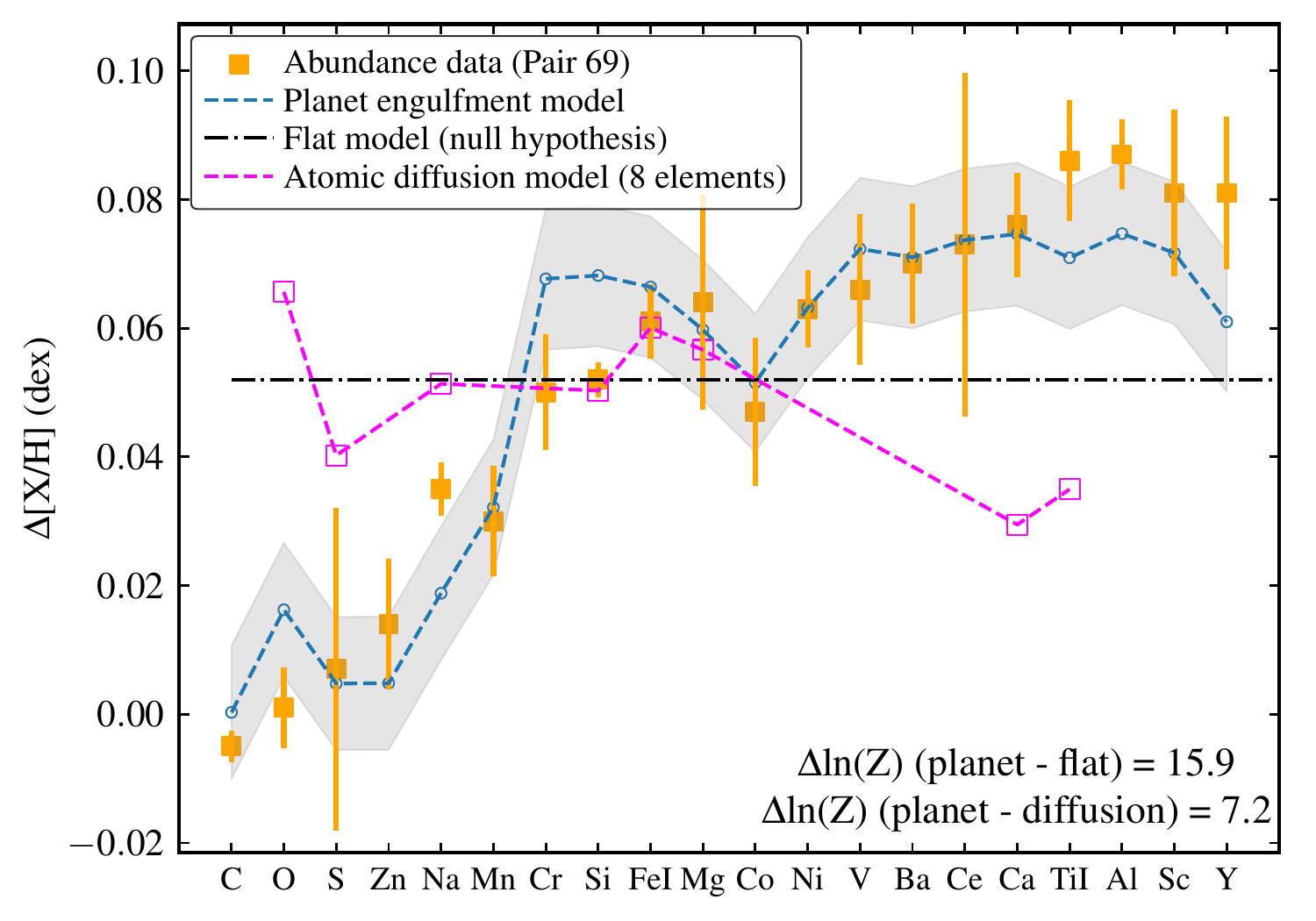} \\
    \subfigimg[width=0.9\hsize]{\textbf{b}}{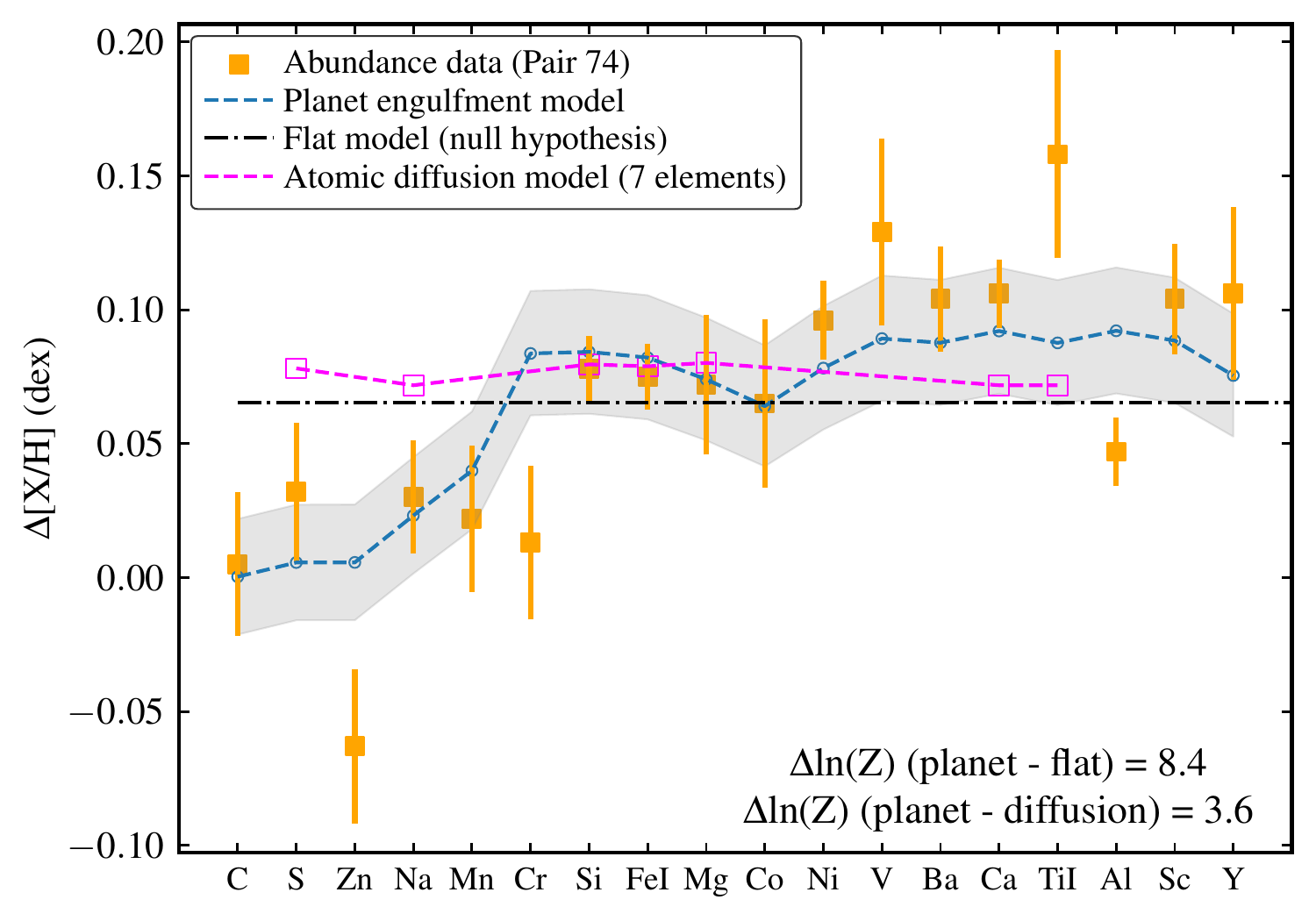} \\
\end{tabular}
\caption{\textbf{Abundance differences ($\Delta$[X/H]) in our candidate pairs.} Same as Figure \ref{fig:fitting_data}\,\textbf{a}, but for: \textbf{a.} Pair 69; and \textbf{b.} Pair 74. The error bars are 1\,$\sigma$ uncertainties of the observed abundances.}
\label{fig:fitting_A5}
\end{figure}

\begin{figure}[hbp]
\centering
\begin{tabular}{c}
    \subfigimg[width=0.9\hsize]{\textbf{a}}{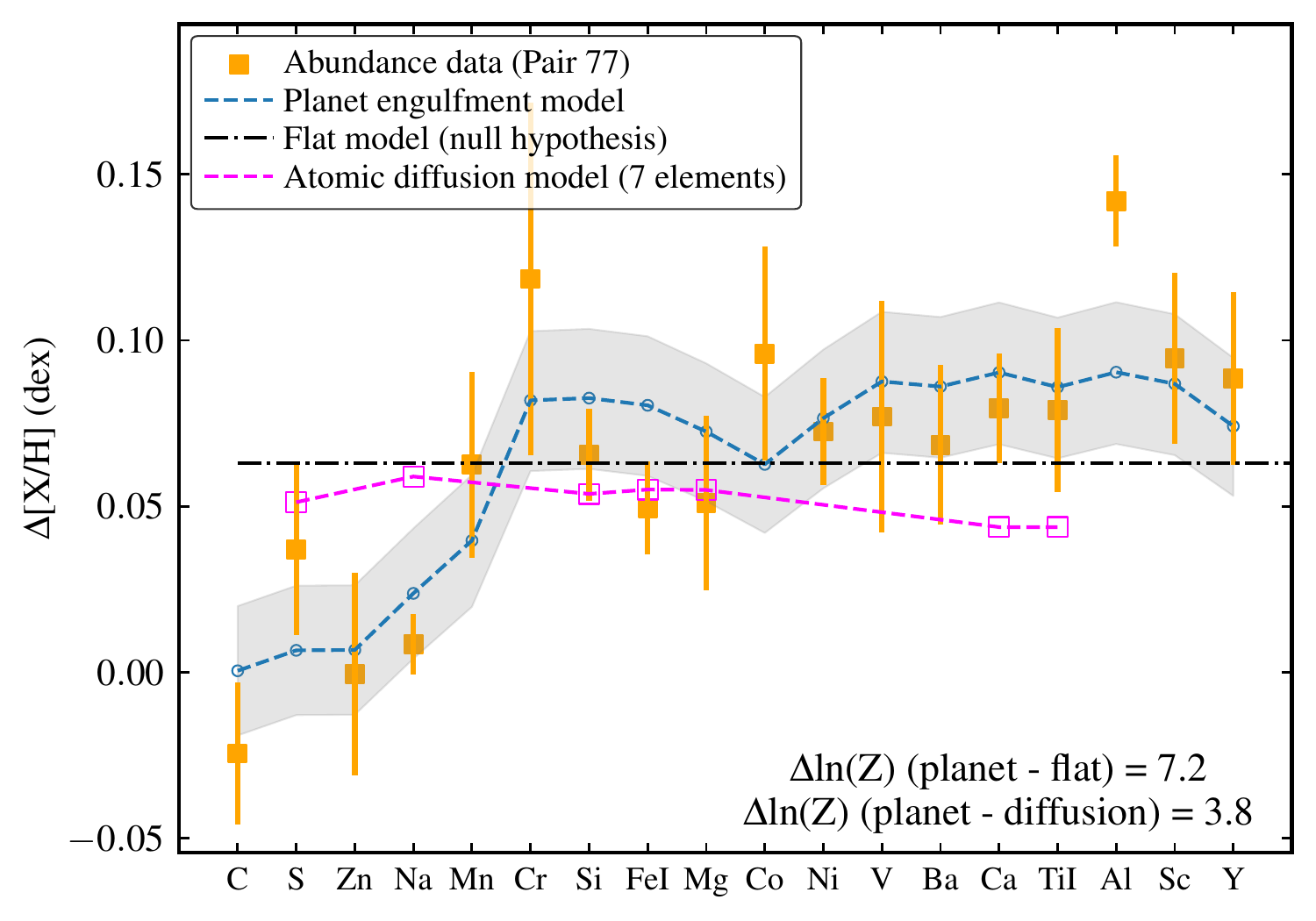} \\
    \subfigimg[width=0.9\hsize]{\textbf{b}}{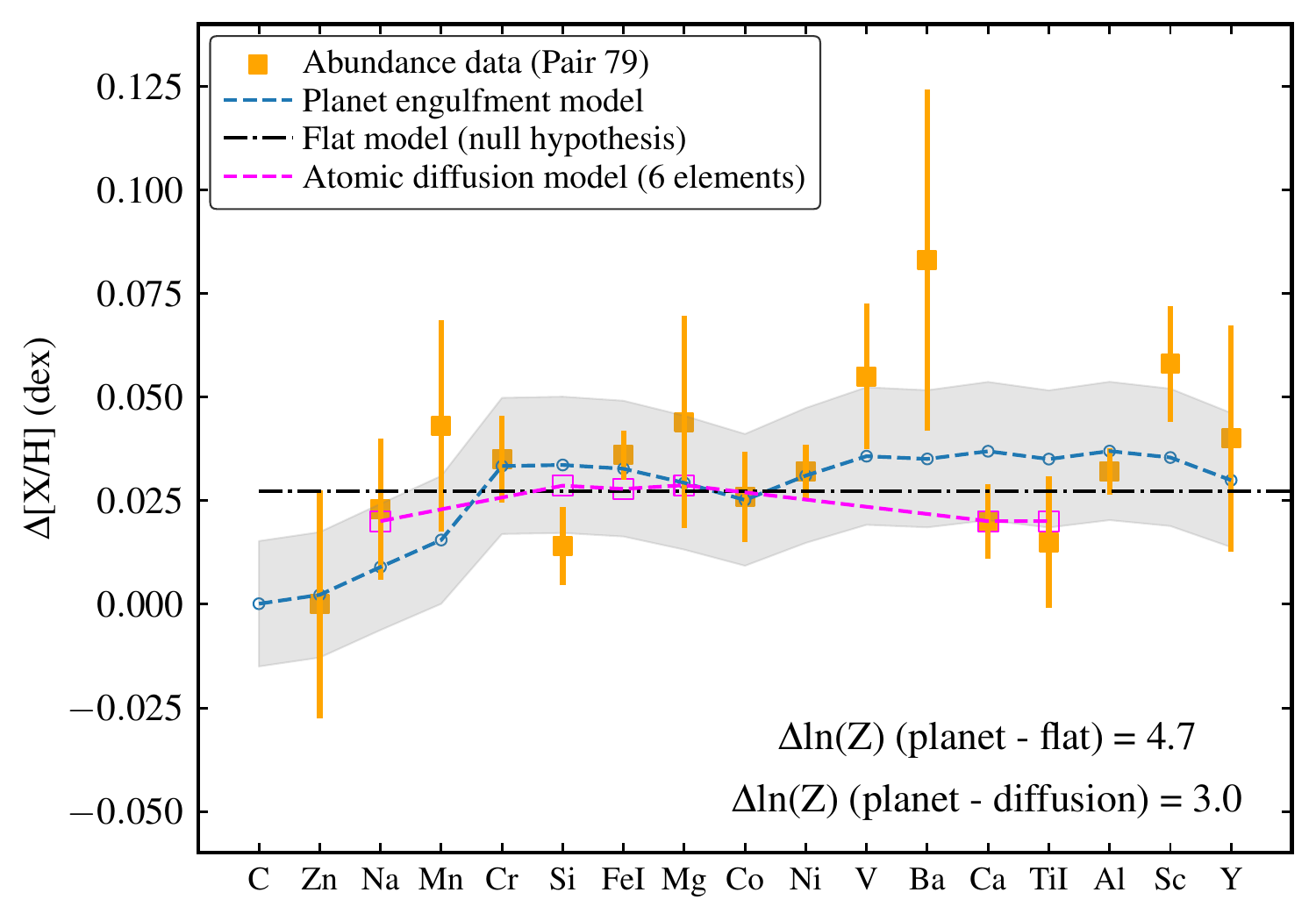} \\
\end{tabular}
\caption{\textbf{Abundance differences ($\Delta$[X/H]) in our candidate pairs.} Same as Figure \ref{fig:fitting_data}\,\textbf{a}, but for: \textbf{a.} Pair 77; and \textbf{b.} Pair 79. The error bars are 1\,$\sigma$ uncertainties of the observed abundances.}
\label{fig:fitting_A6}
\end{figure}

\begin{figure}[hbp]
\centering
\begin{tabular}{c}
    \subfigimg[width=0.9\hsize]{\textbf{a}}{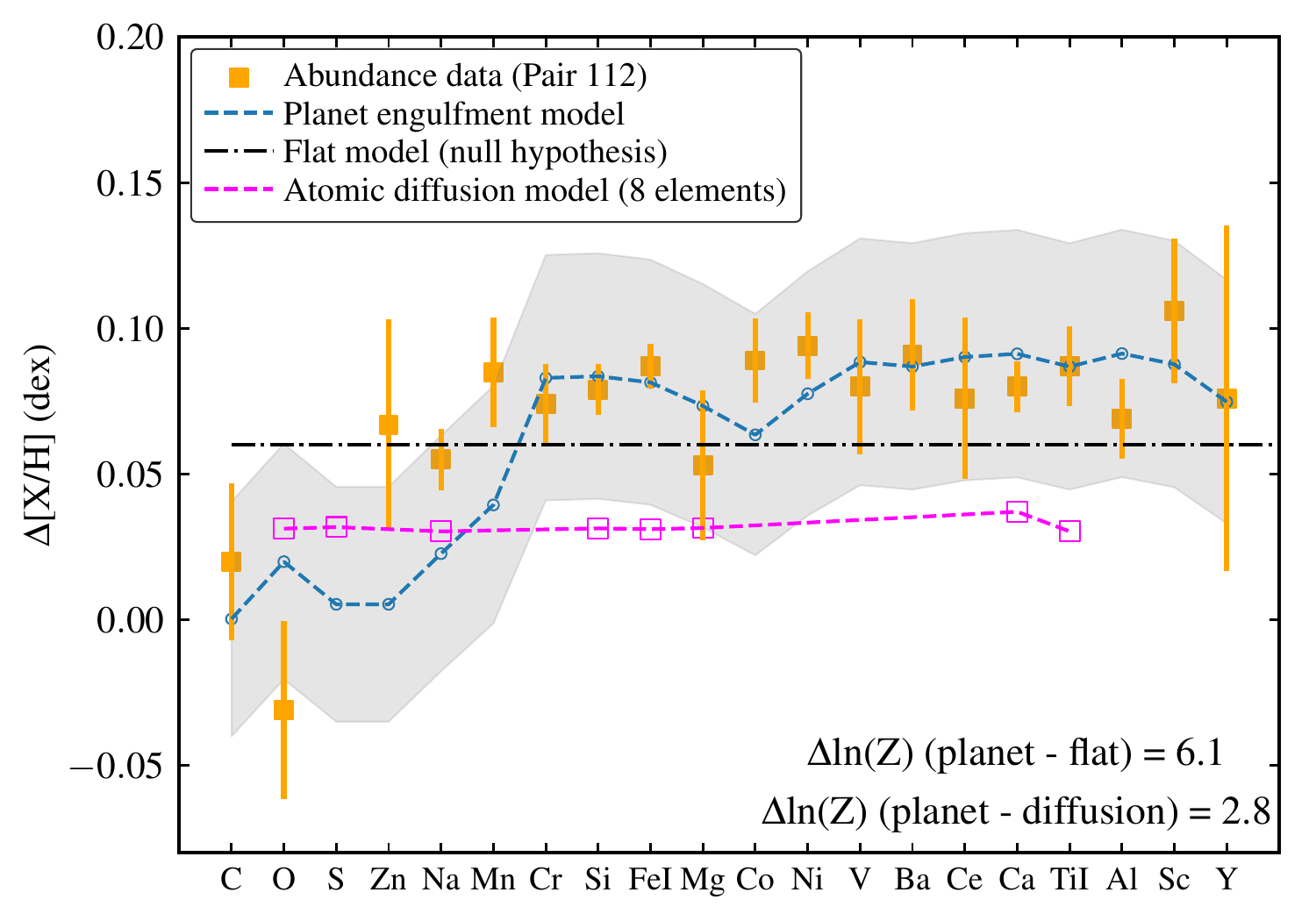} \\
    \subfigimg[width=0.9\hsize]{\textbf{b}}{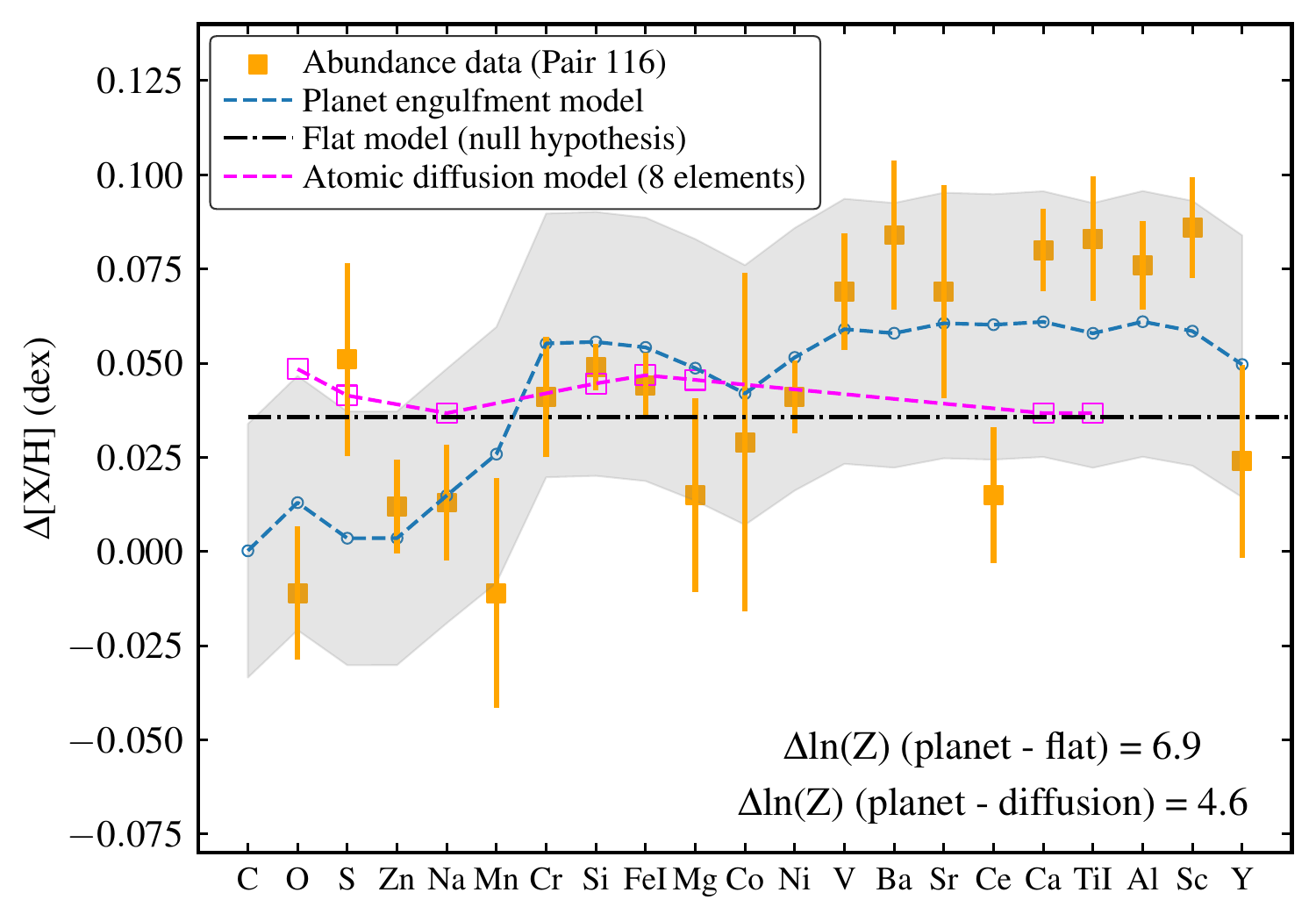} \\
\end{tabular}
\caption{\textbf{Abundance differences ($\Delta$[X/H]) in our candidate pairs.} Same as Figure \ref{fig:fitting_data}\,\textbf{a}, but for: \textbf{a.} Pair 112; and \textbf{b.} Pair 116. The error bars are 1\,$\sigma$ uncertainties of the observed abundances.}
\label{fig:fitting_A7}
\end{figure}

\setcounter{table}{0}
\renewcommand{\tablename}{Extended Table}

\begin{table}
\begin{minipage}{190mm}
\caption{\textbf{Fitting results of seven co-moving pairs of stars exhibiting distinct chemical signatures of ingestion of planetary ingestion.}} 
\centering
\label{table:fitting_results}
\begin{tabular}{@{}ccccrrr@{}}
\hline
Pair ID & $|$Slope$|^a$ & $\sigma_{|{\rm Slope}|}$ & $M_{\rm E}$ & ln(Z)$_{\rm planet}$ & $\Delta$ln(Z)$^b$ & $\Delta$ln(Z)$_{\rm atom}$$^c$ \\ 
 & ($10^{-5}$\,K$^{-1}$) & ($10^{-5}$\,K$^{-1}$) & \mearth & & & \\
\hline
 69   &   5.08  &  0.21  &  3.07  &   49.4  &  15.9  &  7.2 \\  
 74   &   7.16  &  1.39  &  4.80  &   29.0  &   8.4  &  3.6 \\  
 77   &   9.92  &  1.17  &  1.74  &   30.7  &   7.2  &  3.8 \\  
 79   &   5.14  &  1.18  &  3.16  &   28.9  &   4.7  &  3.0 \\  
112   &   7.53  &  1.14  &  8.34  &   26.4  &   6.1  &  2.8 \\  
116   &   9.71  &  0.65  &  5.81  &   33.2  &   6.9  &  4.6 \\  
124   &  13.85  &  0.75  &  3.07  &   27.2  &  12.3  &  5.7 \\  
\hline
\end{tabular}
{\raggedright $^a$ Absolute slopes of \tcond\ trends (fittings to all elements). \par}
{\raggedright $^b$ For planetary ingestion model against flat model (null hypothesis). \par}
{\raggedright $^c$ For planetary ingestion model against atomic diffusion model (for available elements). \par}
\end{minipage}
\end{table}

\end{document}